\documentclass[twocolumn,showpacs,preprintnumbers,amsmath,prd,amssymb,superscriptaddress]{revtex4}

\pdfoutput=1

\usepackage{graphicx}
\usepackage{dcolumn}
\usepackage{bm}
\usepackage{multirow}
\usepackage{float}
\usepackage{color,soul}
\usepackage{hyperref}
\usepackage{makecell}
\usepackage{tabularx} 
\usepackage{url}

\def\th232{\rm{ ^{232} Th }}
\def\u238{\rm{ ^{238} U }}
\def\ur235{\rm{ ^{235} U }}
\def\halflife{\tau_{1/2}}

\begin{document}

\title{Neutron background measurements with a hybrid neutron detector at
the Kuo-Sheng Reactor Neutrino Laboratory}

%

\newcommand{\deu}{Department of Physics,
Dokuz Eyl\"{u}l University, Buca, \.{I}zmir TR35390, Turkey}
\newcommand{\metu}{Department of Physics,
Middle East Technical University, Ankara TR06800, Turkey}
\newcommand{\as}{Institute of Physics, Academia Sinica, Taipei 115, Taiwan}
\newcommand{\thu}{Department of Engineering Physics, Tsinghua University,
Beijing 100084, China}
\newcommand{\ihep}{Institute of High Energy Physics, Chinese
Academy of Science, Beijing 100039, China}
\newcommand{\ciae}{Department of Nuclear Physics,
Institute of Atomic Energy, Beijing 102413, China}
\newcommand{\bhu}{Department of Physics, Institute of Science,
Banaras Hindu University, Varanasi 221005, India}
\newcommand{\nku}{Department of Physics, Nankai University,
Tianjin 300071, China}
\newcommand{\scu}{College of Physical Science and Technology,
Sichuan University, Chengdu 610064, China}
\newcommand{\kiran}{Physics Department, KL University, Guntur 522502, India}
\newcommand{\corr}{htwong@phys.sinica.edu.tw}
\newcommand{\corrmd}{muhammed.deniz@deu.edu.tr}

\author{ A.~Sonay }  \affiliation{ \deu } \affiliation{ \as}
\author{ M.~Deniz } \altaffiliation[Corresponding Author: ]{ \corrmd }
\affiliation{ \deu } \affiliation{ \as }
\author{ H.~T.~Wong } \altaffiliation[Corresponding Author: ]{ \corr } \affiliation{ \as }
\author{ M.~Agartioglu }  \affiliation{ \deu } \affiliation{ \as }
\author{ G.~Asryan }  \affiliation{ \as }
\author{ J.~H.~Chen }  \affiliation{ \as }
\author{ S.~Kerman } \affiliation{ \deu } \affiliation{ \as }
\author{ H.~B.~Li }  \affiliation{ \as }
\author{ J.~Li }  \affiliation{ \ihep }
\author{ F.~K.~Lin }  \affiliation{ \as }
\author{ S.~T.~Lin }  \affiliation{ \as } \affiliation{ \scu }
\author{ B.~Sevda } \affiliation{ \deu } \affiliation{ \as }
\author{ V.~Sharma }  \affiliation{ \as } \affiliation{ \bhu }
\author{ L.~Singh }  \affiliation{ \as } \affiliation{ \bhu }
\author{ M.~K.~Singh } \affiliation{ \as } \affiliation{ \bhu }
\author{ V.~Singh }  \affiliation{ \bhu }
\author{ A.~K.~Soma }  \affiliation{ \as } \affiliation{ \bhu }
\author{ S.~W.~Yang }  \affiliation{ \as }
\author{ Q.~Yue }  \affiliation{ \thu }
\author{ I.~O.~Y{\i}ld{\i}r{\i}m } \affiliation{ \metu }
\author{ M.~Zeyrek } \affiliation{ \metu }
\collaboration{The TEXONO Collaboration}

\date{\today}



\begin{abstract}

We report {\it in situ} neutron background measurements at the
Kuo-Sheng Reactor Neutrino Laboratory (KSNL) by a hybrid neutron
detector (HND) with a data size of 33.8 days under identical
shielding configurations as during the neutrino physics data taking.
The HND consists of BC-501A liquid and BC-702 phosphor powder
scintillation neutron detectors, which is sensitive to both fast
and thermal neutrons, respectively. Neutron-induced events for
the two channels are identified and differentiated by pulse shape
analysis, such that background of both are simultaneously measured.
The fast neutron fluxes are derived by an iterative unfolding algorithm.
Neutron induced background in the germanium detector under the same fluxes,
both due to cosmic-rays and ambient radioactivity, are derived and compared
with the measurements. The results are valuable to background understanding
of the neutrino data at the KSNL. In particular, neutron-induced background
events due to ambient radioactivity as well as from reactor operation are
negligible compared to intrinsic cosmogenic activity and ambient $\gamma$-activity.
The detector concept and analysis procedures are applicable to neutron background
characterization in similar rare-event experiments.

\end{abstract}

\maketitle

\section{Introduction}

The TEXONO Collaboration~\cite{huni} is
pursuing experimental investigation of
neutrino physics~\cite{hbli_prl,texononue,jwchen,hjphys},
as well as weakly interacting massive particle (WIMP)
dark matter~\cite{stlin09}, axions~\cite{chang07} and
other physics searches beyond-standard-model (BSM)~\cite{texononsi}
at the Kuo-Sheng Reactor Neutrino Laboratory (KSNL).
Quantitative understanding of neutron-induced background
and nature of their sources is crucial to these studies.

We report in this article {\it in situ} measurement of thermal
($n_\text{thermal}$) and fast ($n_\text{fast}$) neutron background
at KSNL under identical shielding configurations as during the various
physics data taking. A custom-built hybrid neutron detector (HND), whose
characteristics and performances were reported earlier in our previous
publication~\cite{nd_tech}, are used for these measurements.

The paper is structured as follows.
Highlights of the laboratory KSNL
are presented in Section~\ref{sect::ksnl}.
The unique merits of the HND, its features
and the associated pulse shape discrimination (PSD)
techniques are summarized
in Section~\ref{sect::HND}.
Data taking at the KSNL is discussed in Section~\ref{sect::daq}.
Derivation of the internal contamination of HND are discussed
in Section~\ref{sect::intcontam}.
Results on the measured neutron-induced background in HND,
the calculated neutron fluxes as well as the projected background
to high-purity germanium detectors (HPGe) at  the same location
are presented in Section~\ref{sect::nbkg}.

\section{The Kuo-Sheng Reactor Neutrino Laboratory}
\label{sect::ksnl}

\begin{figure}
\begin{center}
{\bf (a)}\\
\includegraphics[width=8.0cm]{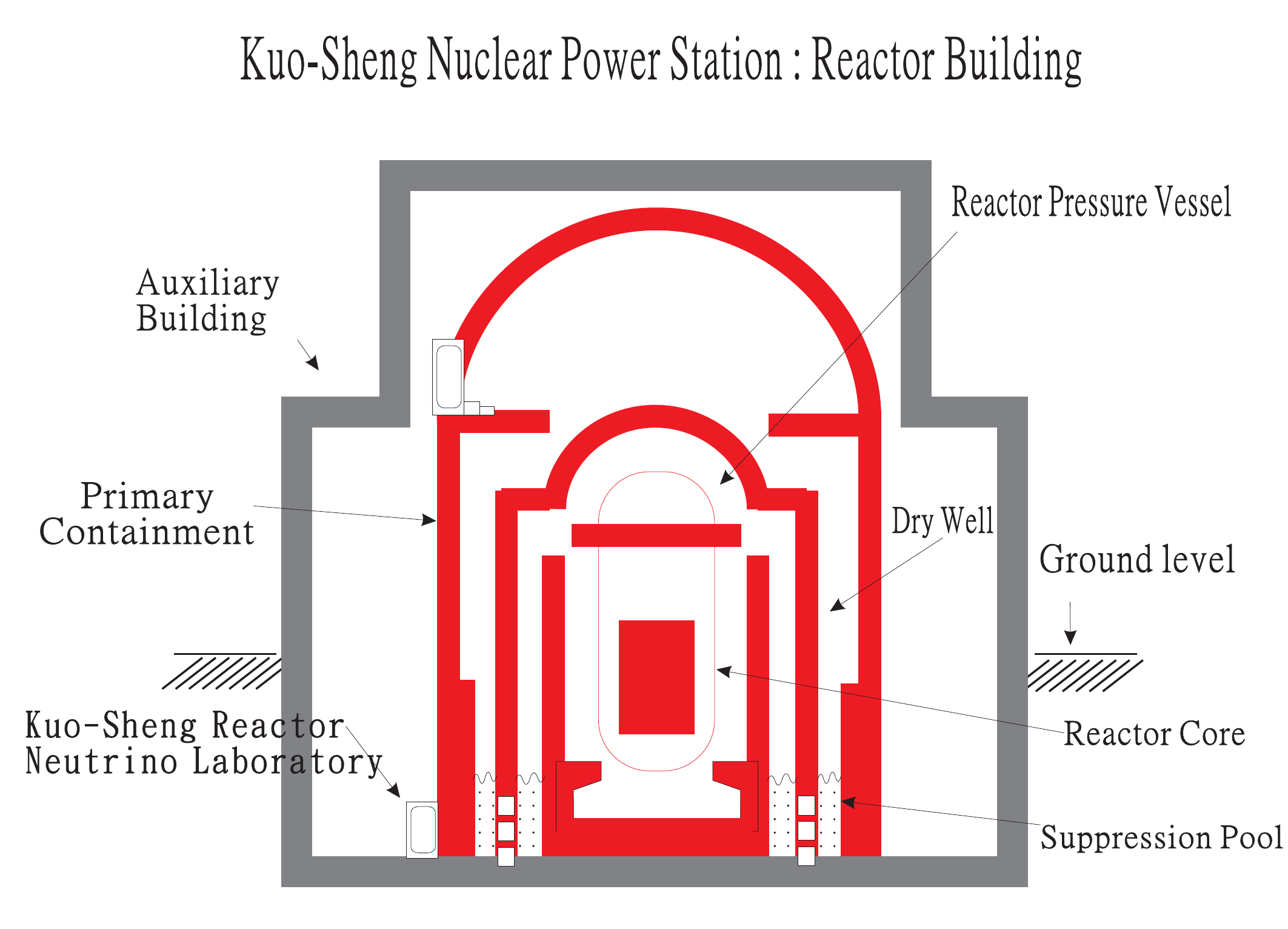}\\
{\bf (b)}\\
\includegraphics[width=8.0cm]{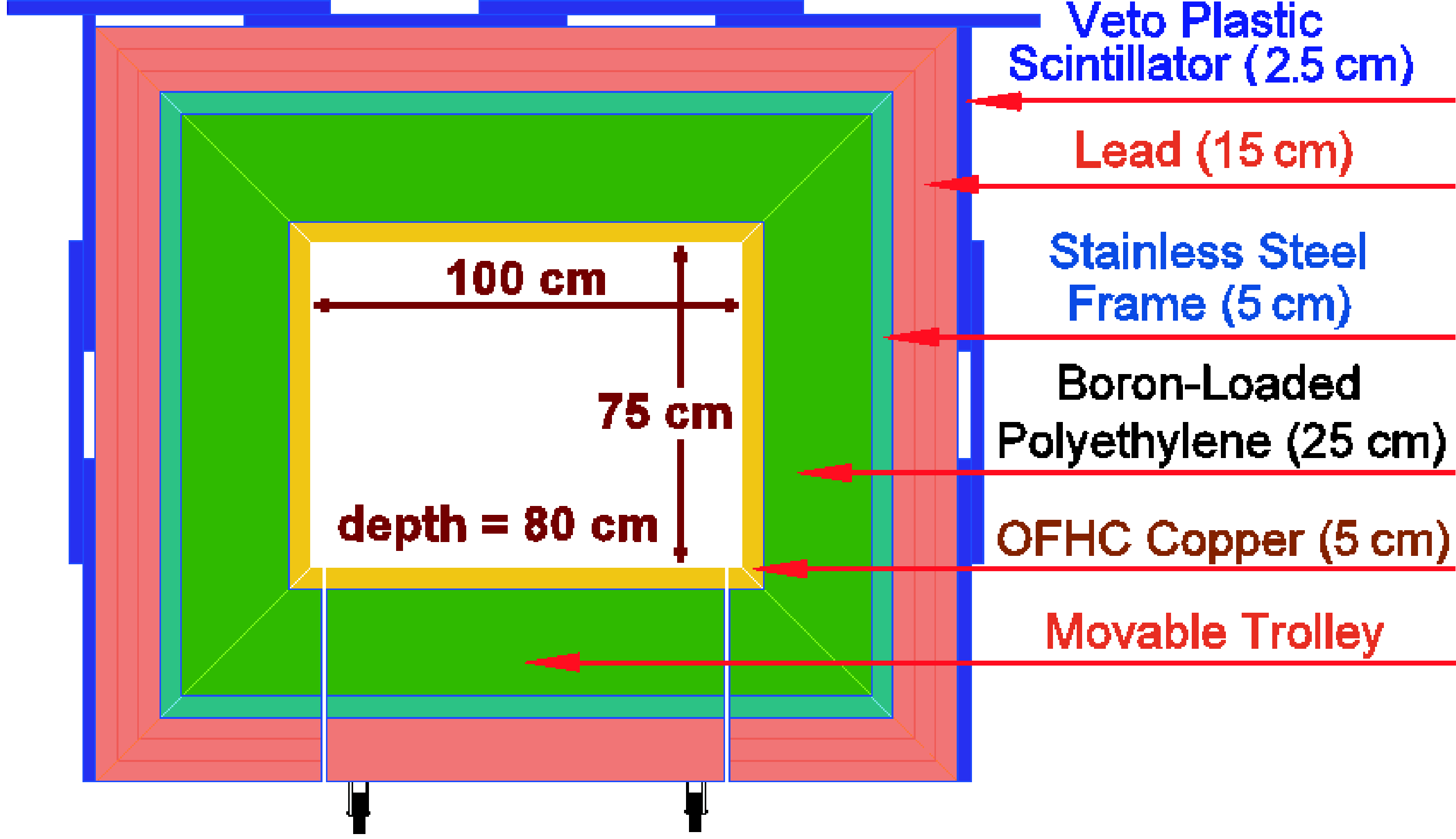}
\caption{
(a) Schematic side view, not drawn to scale,
of the Kuo-Sheng Nuclear Power Station
Reactor Building, indicating the experimental site.
The reactor core-detector distance is about 28~m.
(b) Schematic layout of the general purpose
inner target volume, passive shielding and
cosmic-ray veto panels.
}
\label{fig::ksnlsite}
\end{center}
\end{figure}

The Reactor Neutrino Facility KSNL~\cite{huni,hbli_prl,texononue}
is located at a distance of 28~m from the core \#1
of the Kuo-Sheng Nuclear Power Station
at the northern shore of Taiwan.
The site is at the ground floor of the reactor building
at a depth of 10~m below ground level, with an overburden
of about 30~meter-water-equivalence (mwe).
The nominal thermal power output is 2.9~GW
supplying a $\bar{\nu}_e$-flux of about
$6.4 \times 10^{12} ~ {\rm cm^{-2} s^{-1}}$.
A schematic view is depicted in Figure~\ref{fig::ksnlsite}a.

A multi-purpose ``inner target'' detector volume of
100~cm$\times$80~cm$\times$75~cm is
enclosed by 4$\pi$ passive shielding materials
which have a total weight of about 50 tons.
The shielding provides attenuation
to the ambient neutron and gamma background, and
consists of, from inside out,
5~cm of OFHC copper, 25~cm of boron-loaded
polyethylene, 5~cm of steel, 15~cm of lead,
and cosmic-ray veto scintillator panels.
The schematic layout of the shielding
structure is shown in Figure~\ref{fig::ksnlsite}b.
Different detectors are placed in the
inner volume for the different scientific programs.

The primary cosmic-ray hadronic components
are greatly attenuated by matter
(nuclear interaction length of rock is about 38~cm).
Their fluxes at a shallow depth of $\sim$30~mwe
are therefore negligible.
The neutron background are mostly due to:
(i) cosmic-ray muon-induced interactions~\cite{mu-induced-bkg}  and
(ii) ambient radioactivity followed by $( \alpha , {\rm n} )$ processes
from the materials in the vicinity of the detectors.
The neutron fluxes and their spectra, therefore, depend on the
details of the experimental hardware and shielding configurations,
in addition to the depth. Neutron background measurements
at shallow depth sites have been made~\cite{nbkg-shallow}.
The typical levels for neutrons fluxes above keV are
${\rm \mathcal{O}( 10^{-4},~10^{-3},~ 10^{-5} )~ cm^{-2} s^{-1}}$
for the unshielded, lead-shielded and moderator-shielded
configurations, respectively.

The KSNL shielding structures as shown in Figure~\ref{fig::ksnlsite}b
can attenuate thermal and 1~MeV neutrons by
factors of $\ll$$10^{-6}$ and $\sim$$10^{-4}$,
respectively, according to the simulation studies.
Therefore the ambient unshielded neutron fluxes
are not of relevance to the physics background.
Their direct measurements would be challenging
due to the dominating $\gamma$-background.
The background neutrons are cosmic-ray induced
or originated from radioactivity of hardware
components in vicinity of the detectors.
Measurements of these are the themes of this work,
and will be discussed in details in the subsequent
Sections.

\section{Hybrid neutron detector}
\label{sect::HND}

The design, characteristics and performance of the
HND adopted in this measurement were described
in detail in our previous publication~\cite{nd_tech}.
The HND is a novel detector concept initiated by this work
and was custom-built for {\it this} particular purpose of
{\it in situ} neutron background measurements
at a localized volume at KSNL.

The HND has unique features not provided by
conventional neutron detectors.
It can perform simultaneous measurement of both
thermal and fast neutron fluxes,
in which the neutron induced events are
identified by PSD, thereby
greatly suppressing the much larger $\gamma$-rays background.
The compact dimensions allows sampling of the
fluxes in a relatively localized volume
and under exact shielding configurations $-$
matching well with the size (${\rm \mathcal{O}( 100 )~cm^3}$)
of HPGe detectors.
Commonly-used detectors like
the Bonner multi-spheres array spectrometer~\cite{bonnersphere}
would occupy too much volume to match the space constraints.
Undoped liquid scintillators~\cite{undoped-liqscin}
are sensitive to fast neutrons but
not thermal ones.
Doped liquid scintillators are sensitive
to both thermal and fast neutrons.
Those with signatures $^{6}$Li(n,$\alpha$)$^3$H~\cite{Li6LiqScin}
or $^{10}$B(n,$\alpha$)$^7$Li~\cite{B10LiqScin} can be made compact.
However, the $\alpha$- and proton-recoils
that characterize thermal and fast neutrons, respectively,
are not distinguishable by PSD.
The thermal neutron signatures as low-energy peaks
can be easily contaminated by $\gamma$-background.
Long-term stability
on the performance of the doped scintillators
may also pose technical problems.
Stability has been achieved
in Gd-doped liquid scintillators~\cite{liqscin-Gd}.
The high-energy (n,$\gamma$) signatures for
thermal neutrons are distinctive.
They have been used in low-level
neutron background measurements at underground
laboratories to sensitivities
as low as $\mathcal{O}( 10^{-9} ) ~{\rm cm^{-2} s^{-1}}$.
However, capturing the $\gamma$-rays
would require a detector volume
much larger than that allowed by this application.

\begin{figure}[hbt]
\begin{center}
\includegraphics[width=8.5cm]{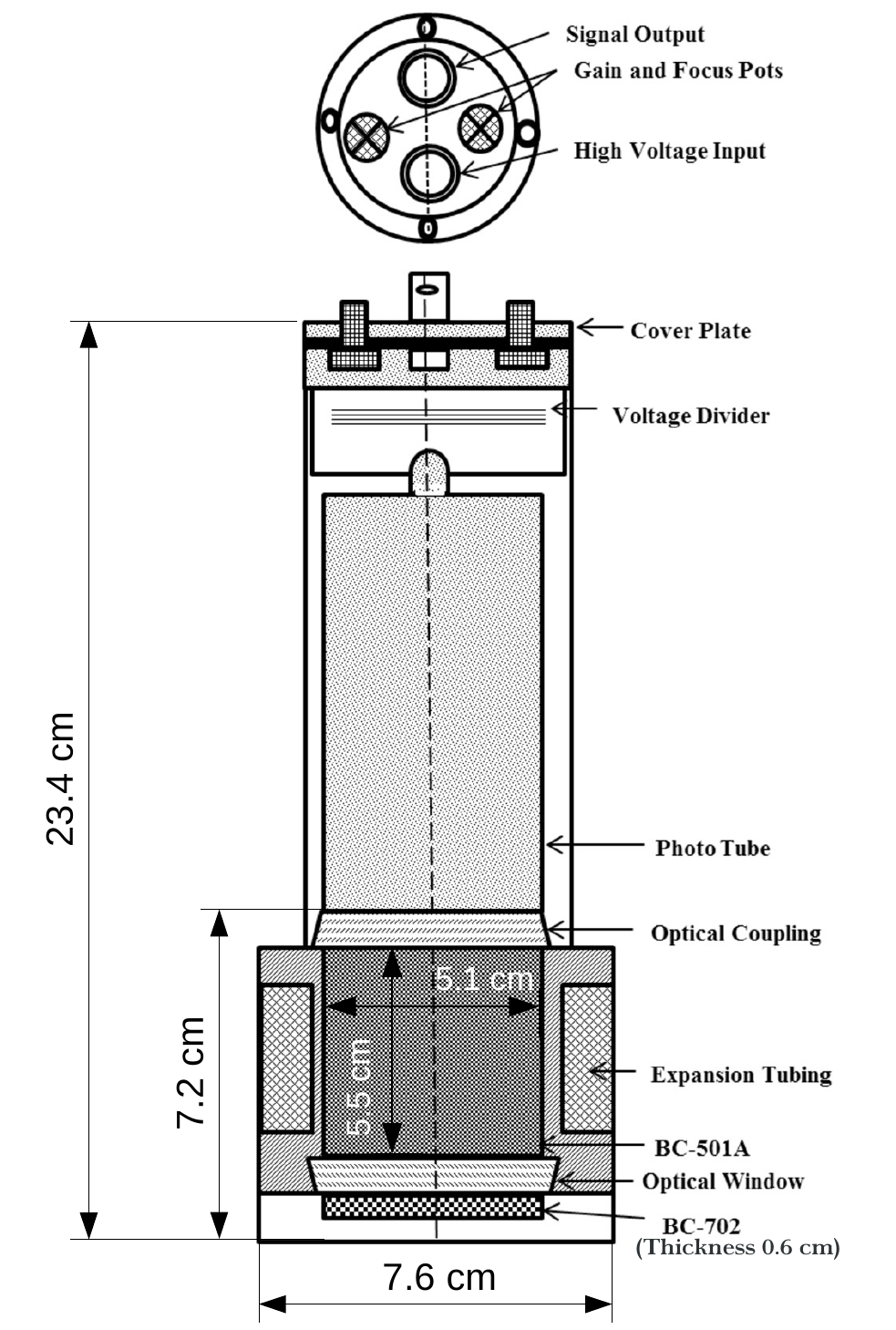}
\caption{
Schematic diagram of the HND.
}
\label{fig::HND}
\end{center}
\end{figure}

\begin{figure}[hbt]
\begin{center}
\includegraphics[width=9cm]{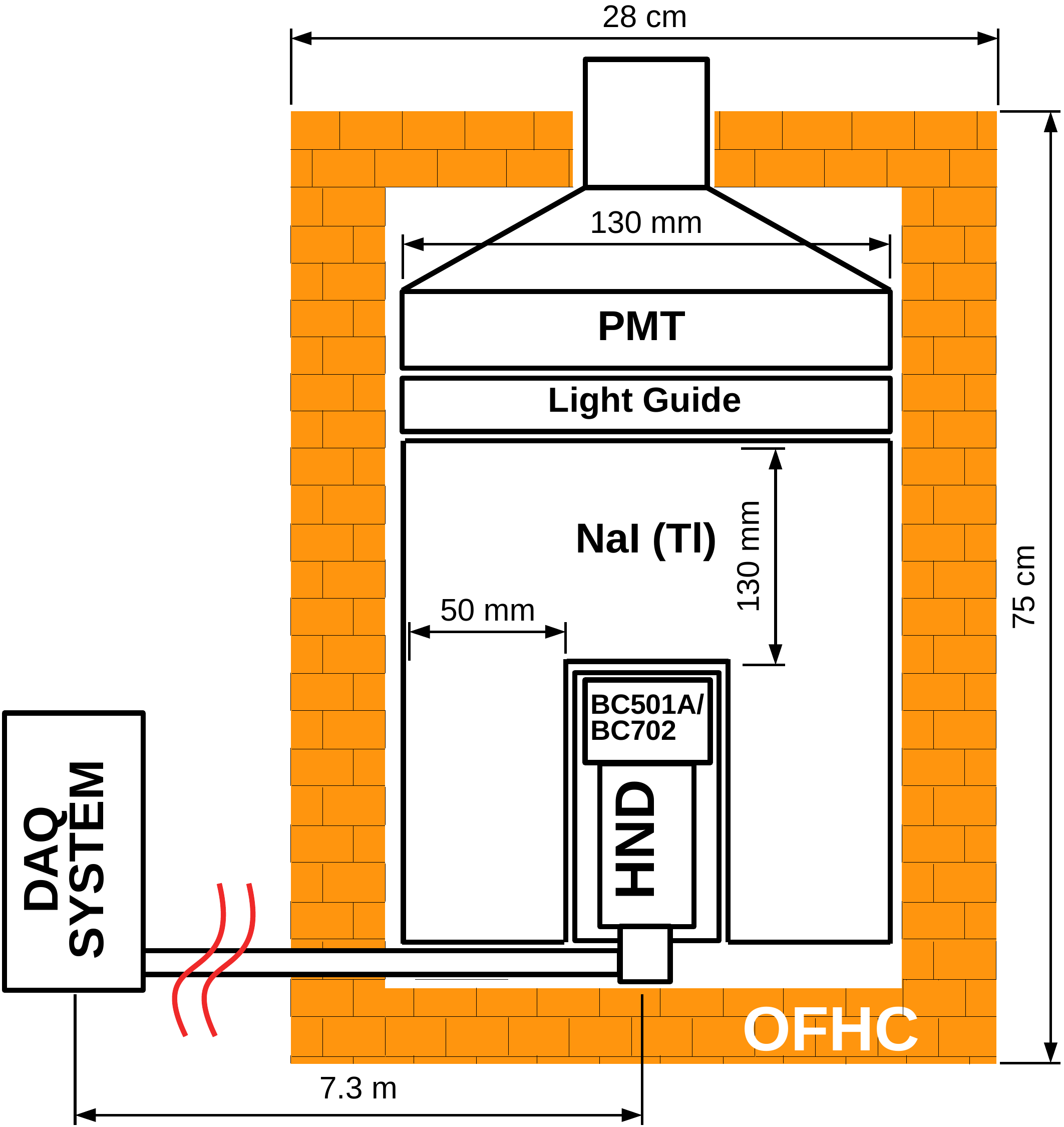}
\caption{
Schematic view of experimental setup inside
the 50 ton shielding structure (not shown).
Signals are brought to the DAQ system via cables
of 7.3 m in length.
}
\label{fig::ksnl_in}
\end{center}
\end{figure}

The HND is constructed with two different target materials$-$
a Bicron BC-501A liquid scintillator with a 0.113~liter cell
volume and a BC-702 scintillator of thickness 0.6 cm enriched
with 95\% $^6$Li as fine ZnS(Ag) phosphor powder $-$ to be read
out by a  5.1 cm diameter photomultiplier (PMT) at the same time.
A schematic drawing of HND is shown in Figure~\ref{fig::HND}.
As depicted in Figure~\ref{fig::ksnl_in}, the HND was installed at
the same location as the various HPGe inside the well of an NaI(Tl)
anti-Compton detector and kept under the same shielding configurations
and data taking conditions. The measured ambient neutron flux is
therefore the same as what the HPGe were exposed to in the physics
data taking.

\begin{figure}[hbt]
\begin{center}
\includegraphics[width=8.5cm]{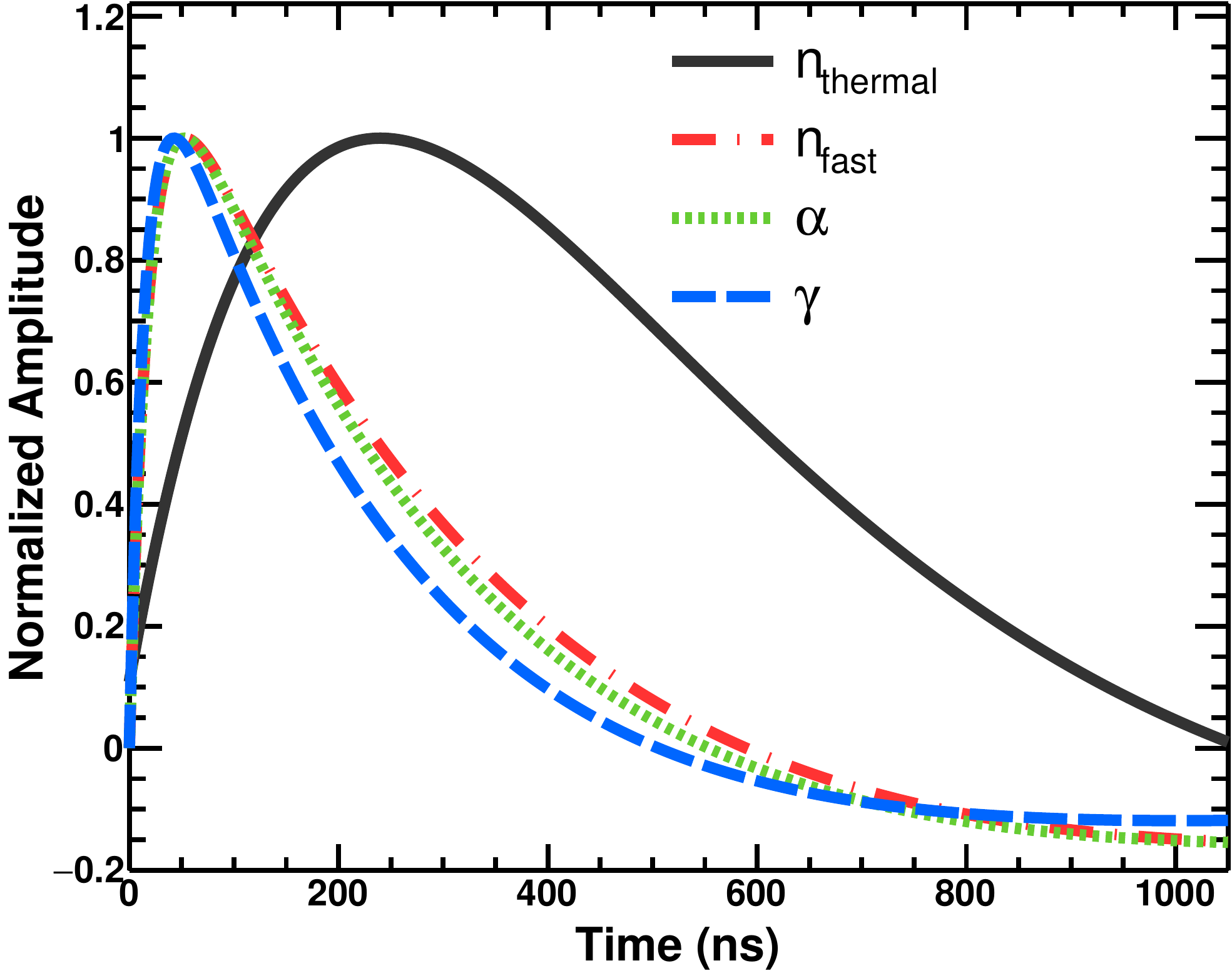}
\caption{
Reference pulse shapes for
$\gamma$-, $n_{\rm fast}$-, and
$n_{\rm thermal}$-induced events from
the HND, from which PSD techniques are
devised to differentiate them.
Pulse shapes of fast neutrons
and alpha-particles are very close
and in practice not distinguishable.
}
\label{fig::pulse_shape}
\end{center}
\end{figure}

Different particles produce different pulse shapes with the
HND~\cite{nd_tech}. The normalized reference pulses of
$\alpha$, $n_{\rm fast}$, $n_{\rm thermal}$ and $\gamma$ are
shown in Figure~\ref{fig::pulse_shape}. In Ref.~\cite{nd_tech},
two independent PSD techniques were developed, which are based
on the parameter of $t_{PSD}$, derived from the ratio of partial
($Q_{p}$) to total ($Q_{t}$) integration of the pulses, and based
on the $B/A$ ratio of individual pulses given as

\begin{eqnarray}
t_{PSD} &=& \frac{Q_{p}}{Q_{t}} = \frac{I[(t_{20}+50 ns):
(t_{20}+150 ns)]}{I[(t_{20}) : (t_{20}+150 ns)]} \nonumber \\
L &=& A\times(e^{-\theta(t-t_{0})}-e^{-\lambda_{s}(t-t_{0})}) \nonumber \\
  &+& B\times(e^{-\theta(t-t_{0})}-e^{-\lambda_{l}(t-t_{0})})~,
\label{eq::psd}
\end{eqnarray}
where $I$ denotes integration of the pulse area and $t_{20}$
represents the time where the pulses reach 20\% of the amplitude.
Here $L$ represents the pulse shape, $A$ and $B$ are the normalization
constants, $t_{0}$ is reference time and $\theta$, $\lambda_{s}$,
and $\lambda_{\ell}$ represent decay constants. Different particles
are identified by different $B/A$ ratios for a specific scintillator.

For this study, a reference pulse is constructed by the superposition
of large number of $\gamma$-ray pulses collected from the 
$^{60}$Co radioactive source. The parameters of decay constants
$\theta$, $\lambda_{s}$, $\lambda_{\ell}$ and reference time $t_{0}$
are obtained from the fitting of the $\gamma$ reference pulse. The
pulse shape is then parameterized as,
\begin{eqnarray}
L &=& A\times\left[(e^{-(t-0.52)/226.6}-e^{-(t-0.52)/17.23}) \right.\nonumber \\
&+& \left.0.115\times(e^{-(t-0.52)/226.6}-1)\right]~,
\label{eq::BA}
\end{eqnarray}
where $t$ is in nanosecond (ns), and $A$ is the only free normalization
parameter that remains to be determined~\cite{nd_tech}. Individual
pulses are fitted with the function given in Eq.~\ref{eq::BA}
in order to identify $\gamma$ against neutron events.

\begin{figure}[hbt]
\begin{center}
{\bf (a)}\\
\includegraphics[width=8.5cm]{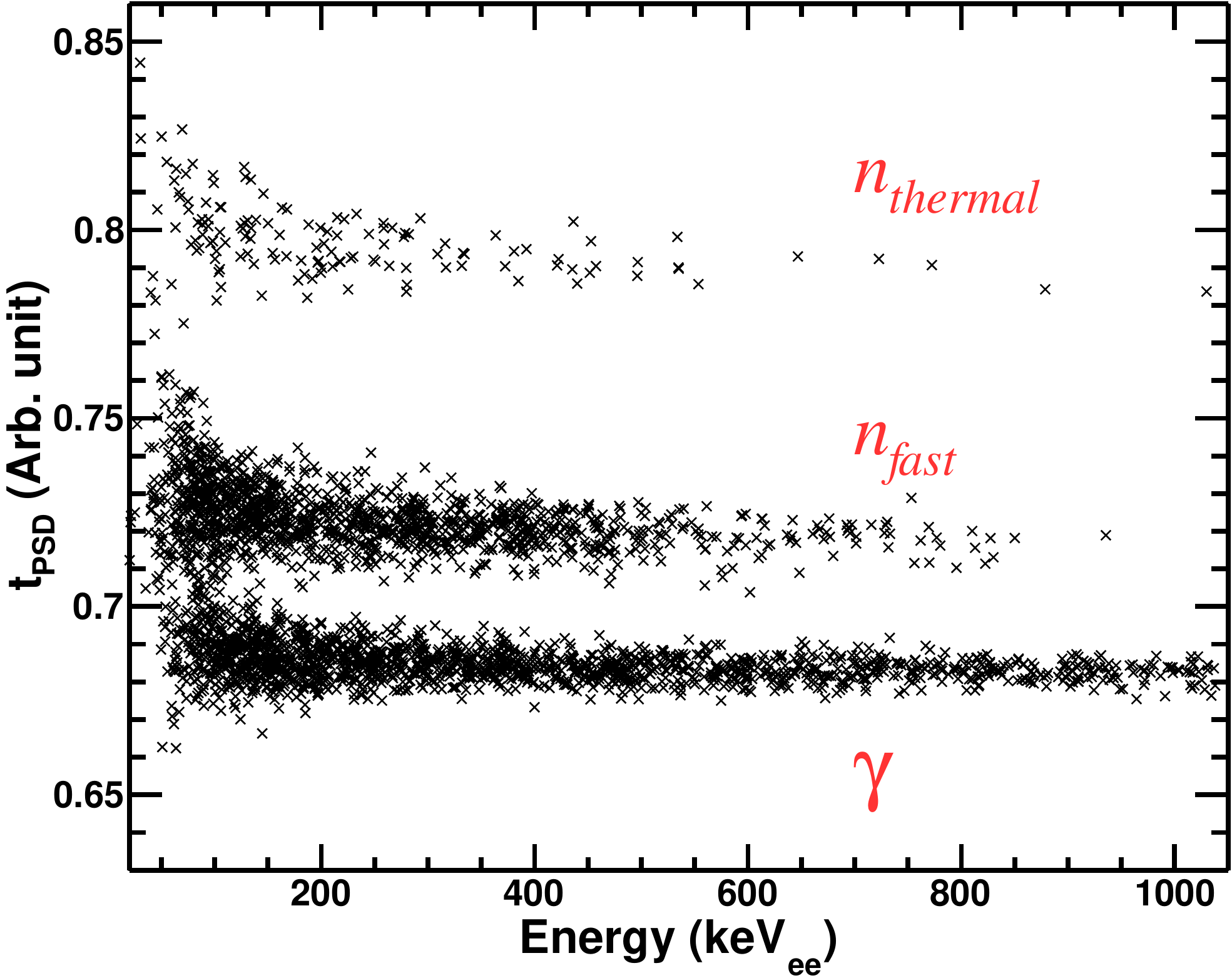} \\
{\bf (b)}\\
\includegraphics[width=8.5cm]{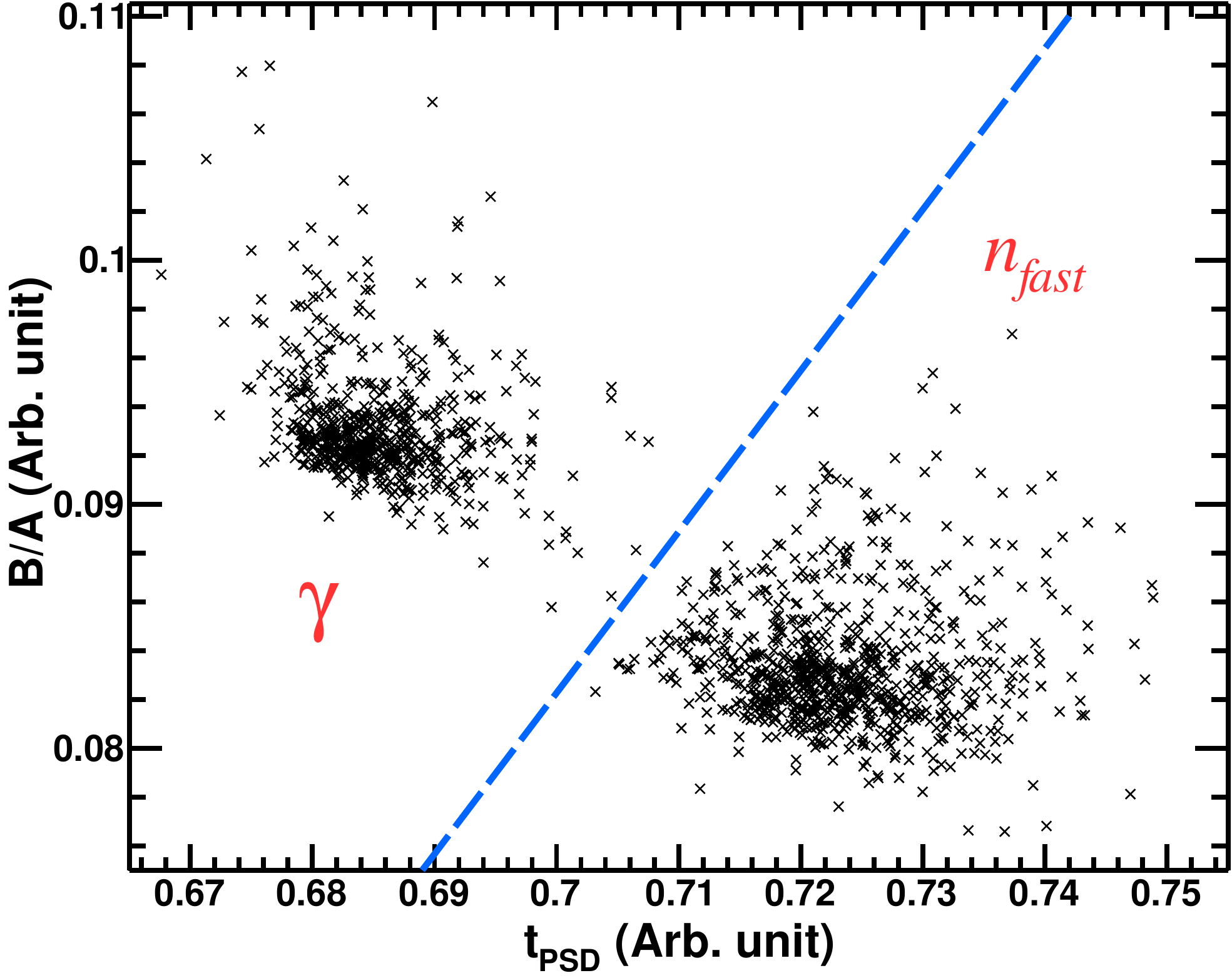}
\caption{
PSD techniques devised to differentiate
$\gamma$- $n_{\rm fast}$- and $n_{\rm thermal}$-induced events,
based on
(a) $t_{PSD}$, and
(b) $B/A$ ratio developed in
Ref.~\cite{nd_tech}.
}
\label{fig::psd_ksnl}
\end{center}
\end{figure}

The $^{241}$AmBe($\alpha$,n) and $^{60}$Co $\gamma-$sources
are used as reference for the $t_{PSD}$ and $B/A$ PSD
techniques, respectively. Adopting the PSD parameters given
in Eq.~\ref{eq::psd} and Eq.~\ref{eq::BA}, three spectral
bands corresponding to $\gamma$, fast and slow neutron
components of the events can be observed, as depicted in
Figure~\ref{fig::psd_ksnl}.

\section{Data taking at Kuo-Sheng neutrino laboratory}
\label{sect::daq}

Several HPGe-based measurements~\cite{hbli_prl,jwchen,stlin09,hjphys,chang07,texononsi}
have been carried out at KSNL. The external dimensions of the HND were selected to resemble those of HPGe. Data were taken at KSNL with the HND placed at the same location as the HPGe~\cite{huni} under identical active and passive shielding configurations, as depicted in Figure~\ref{fig::ksnl_in}. The plastic scintillator panels function as cosmic-ray veto (CR) while the well-shaped NaI(Tl) serves as an anti-Compton (AC) veto detector, and in its cavity the HND (HPGe in early experiments) was placed. The HND+NaI(Tl) detectors were further shielded by oxygen-free high-conductivity (OFHC) copper and placed inside a sealed volume with nitrogen gas flow as a purge of the radioactive radon gas. The setup was installed inside a 50 ton shielding structure~\cite{huni} consisting of, from inside out, OFHC copper, boron-loaded polyethylene, lead and CR panels, for suppression of ambient $\gamma$ and neutron background, and for tagging cosmic-ray induced events.

\begin{figure}[]
\begin{center}
\includegraphics[width=8cm]{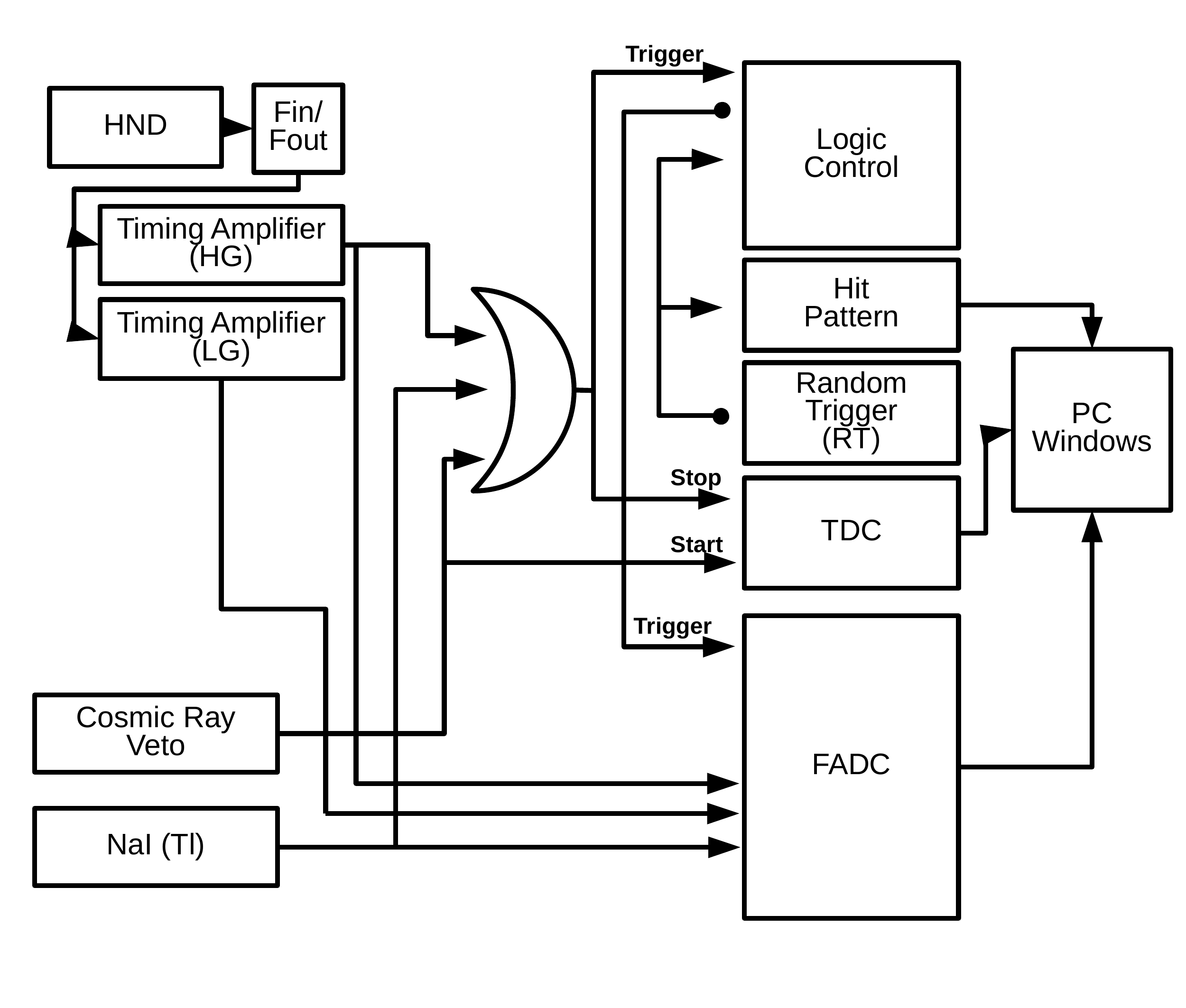}
\caption{Schematic block diagram of the data acquisition system.}
\label{fig::daq}
\end{center}
\end{figure}

The schematic block diagram for the data acquisition (DAQ) system is given in
Figure~\ref{fig::daq}. The HND signals higher than the discriminator threshold
provides the triggers. Signals from other detector components were
recorded to be used for the suppression of AC and CR events in subsequent
offline analysis. The HND signals were processed by two fast timing
amplifiers~\cite{tamp} at different gains and recorded by 8-bit
flash-analog-to-digital converters~\cite{fadc} at 1~GHz sampling rate.
Data taking period lasted more than a month and a total of 33.8 live-time
days of data were collected for subsequent analysis.

The goal of offline analysis is to categorize the events and determine
their respective energy spectra. After standard filtering of events due to
electronic noise and other spurious non-physical triggers, the physical
events are identified as $ \gamma , n_{\rm fast}, n_{\rm thermal} $
from the reference pulse shape information as in Figure~\ref{fig::psd_ksnl}.
The origins of these events are derived from the AC and CR detectors
according to four categories: $\rm{ CR^{\pm} \otimes AC^{\pm}}$ where +(-)
denotes coincidence (anti-coincidence) of the CR or AC with HND. In particular,
the $\rm{ CR^{+} \otimes AC^{-}}$ tag selects CR neutron-induced events,
the $\rm{ CR^{-} \otimes AC^{+}}$ tag is rich in ambient $\gamma$-induced AC
events, while $\rm{ CR^{-} \otimes AC^{-}}$ is the condition for selecting
neutrino- or WIMP-induced candidate events uncorrelated with both CR and AC
systems.

\begin{figure}[hbt]
\begin{center}
\includegraphics[width=8.5cm]{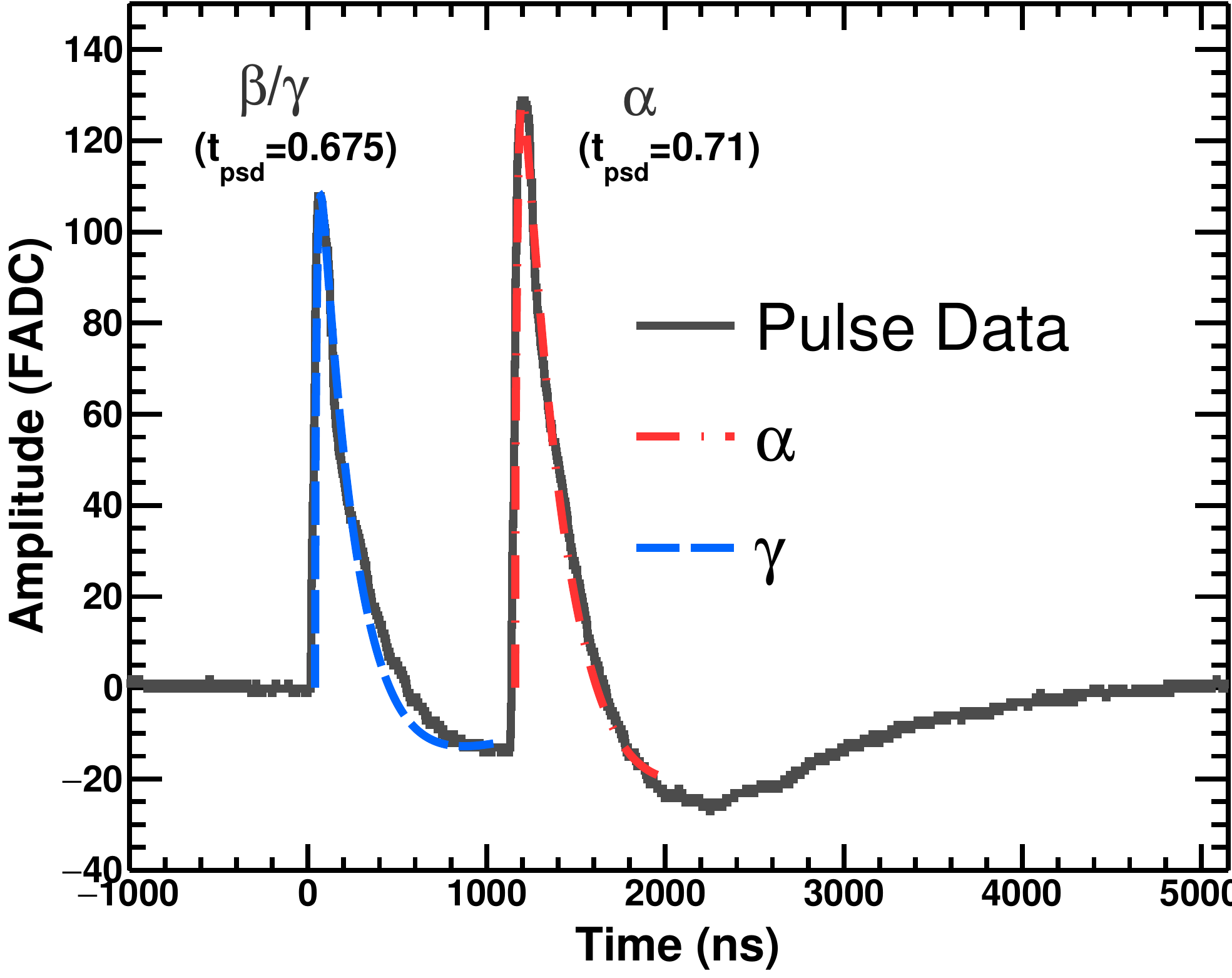}
\caption{
A typical double-pulse event in CR$^{-}$ $\otimes$ AC$^{-}$ channel.}
\label{fig::double_pulse}
\end{center}
\end{figure}

\section{Internal contamination of neutron detector}
\label{sect::intcontam}

The measurement of intrinsic radiopurity of the HND is essential for
determining the ambient neutron background, especially those in
$\rm{ CR^{-} \otimes AC^{-}}$. Nuclear $\alpha$-decays from
the $\u238$ and $\th232$ series can mimic neutron-induced nuclear
recoil signatures, and hence their contributions must be determined.

The PSD characteristics of $\alpha$-events as well as the unique time
correlations of two decay sequences (DS) provide powerful means to
measure contaminations of the $\th232$ and $\u238$ series, from which
the $\alpha$-background can be evaluated, assuming secular equilibrium.

The related DS are~\cite{tori}:
\begin{description}
\item[ \bf DS$_1$: ] Within the $\th232$ series, there is 64\%
branching ratio for $^{212}$Bi to decay via a $\beta$-$\alpha$ cascade $-$
\begin{eqnarray*}
^{212}\rm{Bi} ~ & \rightarrow & ~ ^{212}\rm{Po} ~ + ~ \bar{\nu_e} ~ +
~ e^- ~ + ~ \gamma 's ~  \\
(Q&=&2.25~\rm{MeV} ~ ; ~\halflife =60.6~min) \\
^{212}\rm{Po} ~ & \rightarrow & ~ ^{208}\rm{Pb}~ + ~ \alpha ~
(Q=8.95~\rm{MeV} ~ ; ~ \halflife =0.30~ \mu s)
\end{eqnarray*}
\item[ \bf DS$_2$: ] Within the $\u238$ series, there is
$\alpha$-$\alpha$ cascade from $^{222}$Rn $-$
\begin{eqnarray*}
^{222}\rm{Rn} ~ & \rightarrow & ~ ^{218}\rm{Po} ~ + ~ \alpha ~
(Q=5.59~\rm{MeV} ~ ; ~ \halflife = 3.82~d )  \\
^{218}\rm{Po} ~ & \rightarrow & ~ ^{214}\rm{Pb}~ + ~ \alpha ~
(Q=6.12~\rm{MeV} ~ ; ~ \halflife =3.10~min )
\end{eqnarray*}
\end{description}

Typical example of a double pulse event is displayed in
Figure~\ref{fig::double_pulse}, interpreted as a $\beta - \alpha$
cascade based on PSD. A collection of the delayed pulses in
similar cascades provide the $\alpha$ reference pulse shape
as shown in Figure~\ref{fig::pulse_shape}. The $n_{\rm fast}$/$\gamma$
events are distinguishable while $n_{\rm fast}$/$\alpha$ events are not
distinguishable in an event-by-event basis since the differences in
their pulse shapes are smaller than electronic fluctuations.
The $\alpha$-events of DS$_2$ are mono-energetic and well-separated in time,
and were used to confirm consistency with the resolution and quenching
functions adopted in analysis.

\begin{figure}[hbt]
\begin{center}
{\bf (a)}\\
\includegraphics[width=8.5cm]{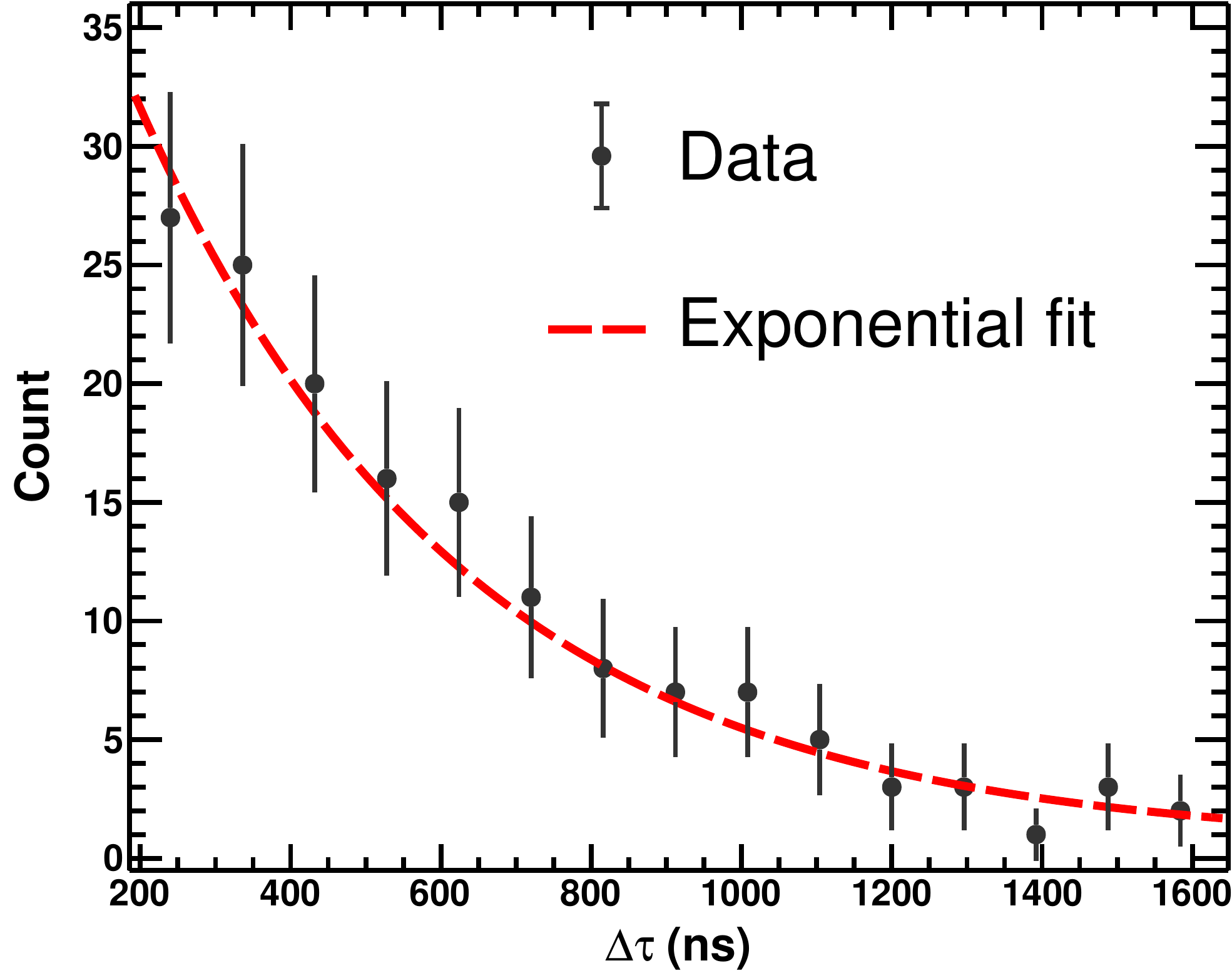} \\
{\bf (b)}\\
\includegraphics[width=8.5cm]{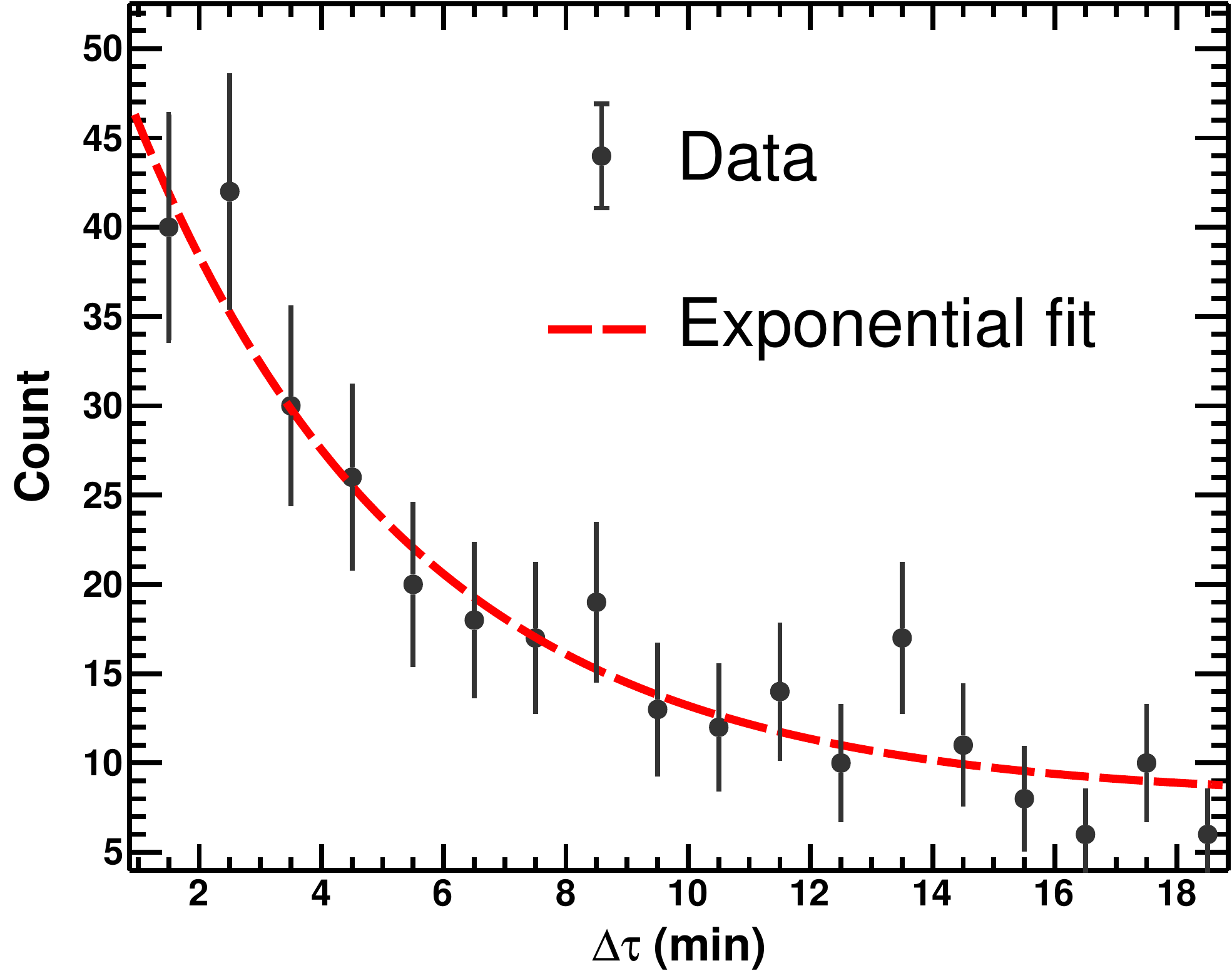}
\caption{
Distribution for
(a) $\beta$-$\alpha$  events from
$^{212}$Bi$\rightarrow^{212}$Po$\rightarrow^{208}$Pb
in $DS_1$,
(b) $\alpha$-$\alpha$ events from
$^{222}$Rn$\rightarrow^{218}$Po$\rightarrow^{214}$Pb
in $DS_{2}$.
}
\label{delaytime}
\end{center}
\end{figure}

The delay-time ($\Delta$t) distributions for the correlated-events
from DS$_{1,2}$ are shown in Figure~\ref{delaytime}(a) and
Figure~\ref{delaytime}(b), respectively. Results of best-fit parameters
to exponential decay functions are displayed in Table~\ref{tab::dsselect}.
%
%

\begin{table}[]
\begin{center}
\caption{
\label{tab::dsselect}
Summary of measured values and inferred radioactivity levels of the
two cascade sequences.}
{\def\arraystretch{1.5}
\begin{tabular}{lcc}
\hline
       & DS$_1$ & DS$_2$ \\
Series & ${}^{232}$Th & ${}^{238}$U \\
\hline \hline
Signatures & $\beta$-$\alpha$ & $\alpha$-$\alpha$ \\
Decays & $\rm{^{212}Bi \rightarrow ^{212}Po }$
       & $\rm{^{222}Rn \rightarrow ^{218}Po}$ \\
       & $\rm{~~~~~~~ \rightarrow ^{208}Pb}$
       & $\rm{~~~~~~~ \rightarrow ^{214}Pb}$\\
$\rm{\chi^2 / n.d.f}$ & 4.7/16 & 9.0/17\\
Half-Life & & \\
~~~~~Nominal & 299~ns & 3.10~min \\
~~~~~Measured & 302 $\pm$ 27~ns & 3.14 $\pm$ 0.39~min \\
\makecell[l]{Counts\\ ~~in 33.8 days} & 366.20 $\pm$ 26.94 & 292.50 $\pm$ 15.43 \\
\makecell[l]{Radioactivity\\ ~~(mBq/kg)} & 0.140 $\pm$ 0.010
                                      & 0.110 $\pm$ 0.006\\
\makecell[l]{Contaminations\\ ~~$\times 10^{-11}$(g/g)}
                     & 2.21 $\pm$ 0.16
                     & 0.89 $\pm$ 0.048\\
\hline
\end{tabular}}
\end{center}
\end{table}

\begin{figure}[hbt]
\begin{center}
\includegraphics[width=8.5cm]{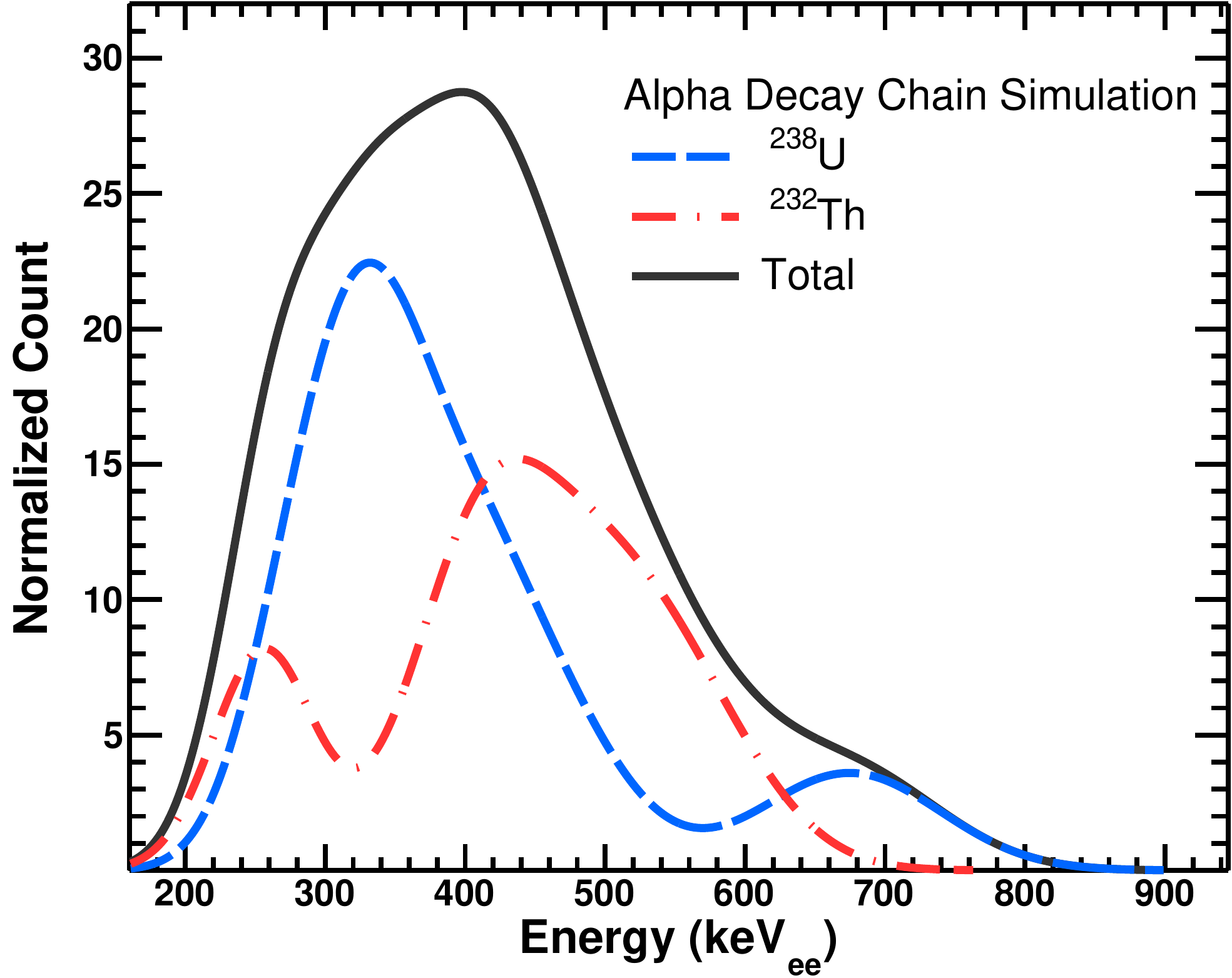}
\caption{
Simulated $\alpha$ energy spectrum of
$\th232$ and $\u238$ decay chains including
detector resolution and quenching effects~\cite{nd_tech}.
Normalization is fixed by the measured cascade sequences
of Table~\ref{tab::dsselect}.
}
\label{fig::parent_alpha_sim}
\end{center}
\end{figure}

The measured event rate of the decay sequences of DS$_{1,2}$
can be used for the estimation of contamination levels of
their long-lived parent isotopes of $\th232$ and $\u238$
in the detector. Simulated $\alpha$ energy spectra of
$\th232$ and $\u238$ series, convoluted with detector
resolution and quenching effects are depicted in
Figure~\ref{fig::parent_alpha_sim}.

The measured half-lives are consistent with nominal values.
The measured event rates can be translated to the radioactivity
and contamination levels of their long-lived parent isotopes of
$\th232$ and $\u238$ in the detector, assuming secular equilibrium.
Simulated $\alpha$ energy spectra of $\th232$ and $\u238$ parent
isotopes convoluted with detector resolution and quenching effects
are depicted in Figure~\ref{fig::parent_alpha_sim}.

\section{Neutron background}
\label{sect::nbkg}

\subsection{Thermal neutron background}

Thermal neutrons are those with kinetic energy below 1~eV and in
thermal equilibrium with the ambient surroundings. Their energy
distribution is described by the Maxwell-Boltzmann distribution
with a most probable energy of $E_{th} \sim {\rm 0.02~eV}$, which
corresponds to a velocity of $v_{th} \sim {\rm 2200 ~ m s^{-1}}$.

The scintillator BC-702 used for thermal neutron measurements does
not provide energy information of the incident neutron. Calculation
of the thermal neutron flux is performed assuming Maxwell-Boltzmann
distributions.

For a neutron flux $\phi_n (E)$  with interaction cross-section
$\sigma(E)$ in the detector, the count rate in the detector
is given by:
\begin{equation}
R_{th}  ~ = ~ N ~ \int ~ \sigma(E) ~ \phi_n (E)dE ~,
\end{equation}
where {\it{N}} is the total number of target nuclei in the detector.
The thermal neutron captured by  $^6{\rm Li}$ in HND
\begin{equation}
n ~ + ~ ^6 Li ~ \rightarrow ~ ^3 H + \alpha
\end{equation}
is inversely proportional to the neutron velocity $v(E)$, such that
\begin{equation}
\sigma(E) ~ = ~ \sigma_{th} ~ \frac{v_{th}}{v(E)} ~,
\end{equation}
where $\sigma_{th} = 940~{\rm b}$.
An isotropic and homogeneous flux distribution
can be described by
\begin{equation}
\phi(E) ~ = ~ v(E) ~ \rho_n (E) ~,
\end{equation}
where $\rho_n (E)$ is the neutron number density at energy $E$
in the detector volume.
The count rate can therefore be expressed as
\begin{equation}
R_{th} ~ = ~ N ~ \sigma_{th} ~ v_{th} ~ \langle \rho_n \rangle ~~,
\end{equation}
where $\langle \rho_n \rangle$ is the energy-averaged
thermal neutron number density.
The average neutron velocity is given by
\begin{equation}
\langle v \rangle ~ =  ~
\frac{\int v(E)~ \rho_n (E)~ dE}{\int \rho_n (E)~ dE}
=\frac{\Phi}{\langle \rho_n \rangle} ~,
\end{equation}
where $\Phi$ is the total flux.
Accordingly, the rate becomes
\begin{equation}
R_{th} ~ = ~ N ~ \sigma_{th} ~ \frac{v_{th}}{\langle v \rangle} ~ \Phi ~.
\end{equation}

\begin{figure}[t]
\begin{center}
{\bf (a)}\\
\includegraphics[width=8.5cm]{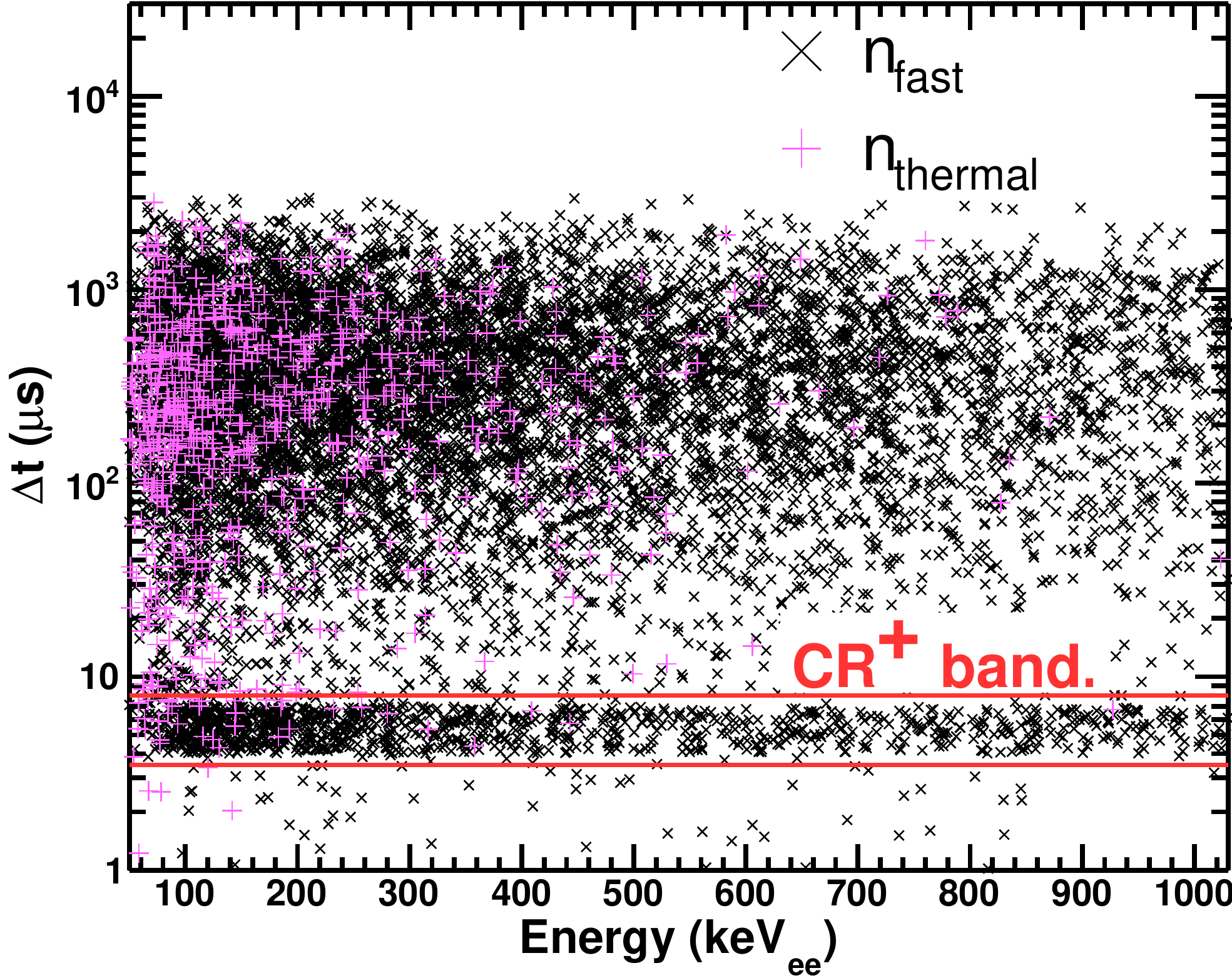} \\
{\bf (b)}\\
\includegraphics[width=8.5cm]{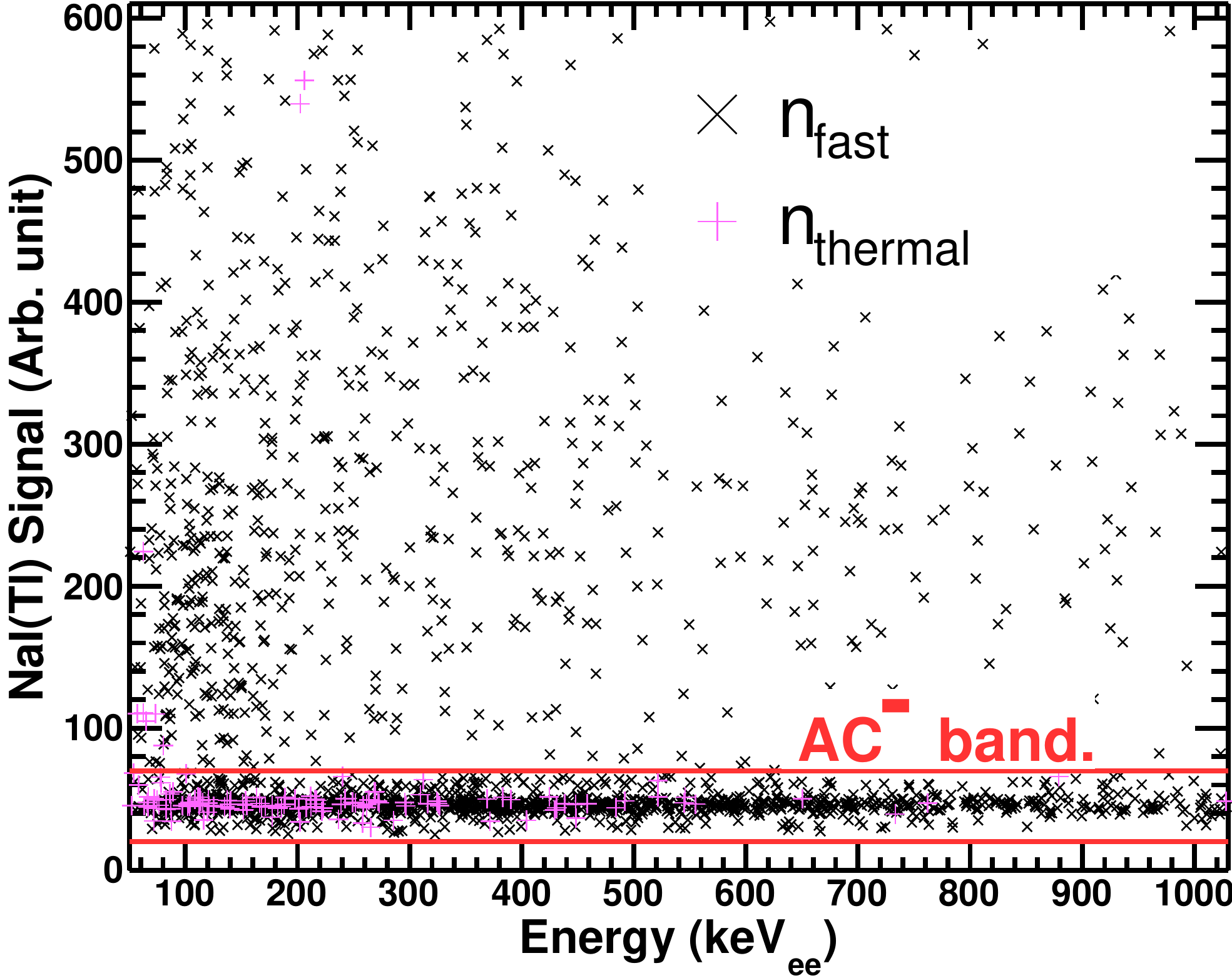}
\caption{Event selection criteria for (a) cosmic-ray CR$^{+}$,
and (b) anti-Compton AC$^{-}$ events, showing most thermal neutrons
are with CR$^{-}$ $\otimes$ AC$^{-}$ tag.}
\label{fig::CRT_ACT}
\end{center}
\end{figure}

Maxwell-Boltzmann distribution for thermal neutrons
gives rise to the relation
\begin{equation}
\frac{\langle v \rangle}{ v_{th} } ~ = ~
\frac{2}{\sqrt{\pi}} ~~.
\end{equation}

Accordingly, the total neutron flux is related
to the measured count rate as
\begin{eqnarray}
\Phi_n ~  = ~
\frac{ 2 R_{th} }{N \sigma_{th} \sqrt{\pi} } .
\end{eqnarray}

The measured thermal neutron rate at KSNL with HND BC-702 is

\begin{eqnarray}
R_{th} = (4.15 \pm 0.12) \times 10^{-4} ~ {\rm counts ~ s^{-1}}~~.
\end{eqnarray}

With a total number of
$N = 1.41 \times 10^{22}$ $^{6}$Li atoms in BC-702 \cite{birk,bicron},
the corresponding total thermal neutron flux is
\begin{equation}
\Phi_n = (3.54 \pm 0.10) \times 10 ^{-5} ~ {\rm cm^{-2}~s^{-1}} ~~.
\end{equation}

\begin{figure}[hbt]
\begin{center}
\includegraphics[width=8.5cm]{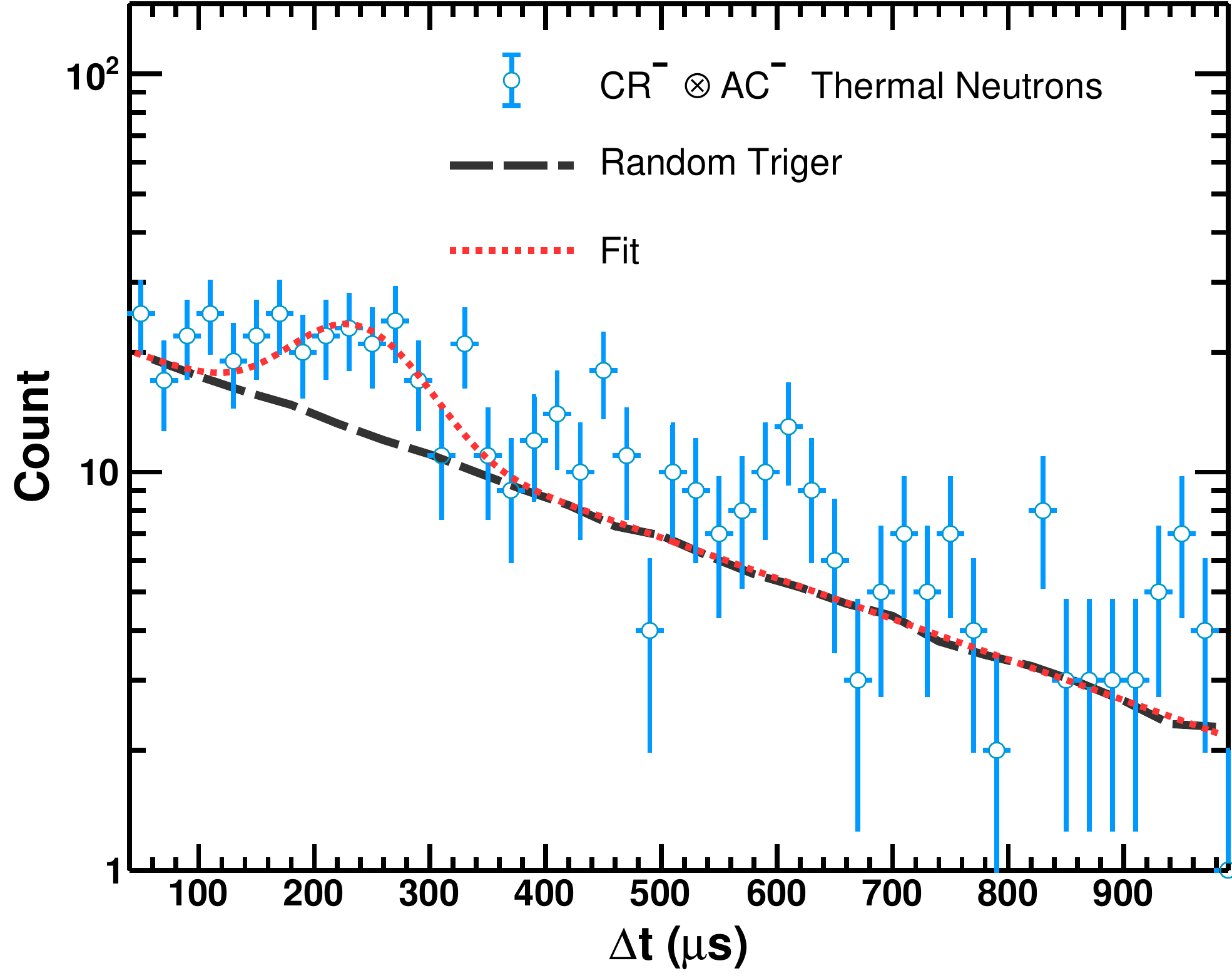} \\
\caption{
The time difference between
CR$^{-}$ $\otimes$ AC$^{-}$ thermal neutron
and the previous cosmic-ray events.
Comparison with random trigger events up to
$\sim$400~$\mu$s indicate that about 20\% of thermal
neutrons can be matched to their parent cosmic-ray events.
}
\label{fig::time_corr}
\end{center}
\end{figure}

The majority of the thermal neutron events are with
CR$^{-}$ $\otimes$ AC$^{-}$ tag uncorrelated with the
other detector systems, as depicted in Figure~\ref{fig::CRT_ACT}.
The time difference between these events
with the previous CR$^+$ tag is displayed in Figure~\ref{fig::time_corr},
in which  accidental coincidence from random trigger events are superimposed.
An excess is observed with a correlation time scale of about 200~$\mu$s,
indicating that part (20\%) of thermal neutron capture events
can be matched to the thermalization of specific cosmic-ray events.
The time scale corresponds to that necessary for the cosmic-induced
high-energy neutrons to lose their energy,
get thermalized and diffuse into the localized BC-702 volume.
Similar distribution profiles were measured and compared with
simulations with gadolinium-loaded liquid scintillator
at a shallow depth~\cite{Dt-mu-ncap}.

\subsection{Measured nuclear recoil spectra; Evaluated fast neutron flux;
Projected HPGe background}

\begin{figure}
\begin{center}
{\bf (a)}\\
\includegraphics[width=8.5cm]{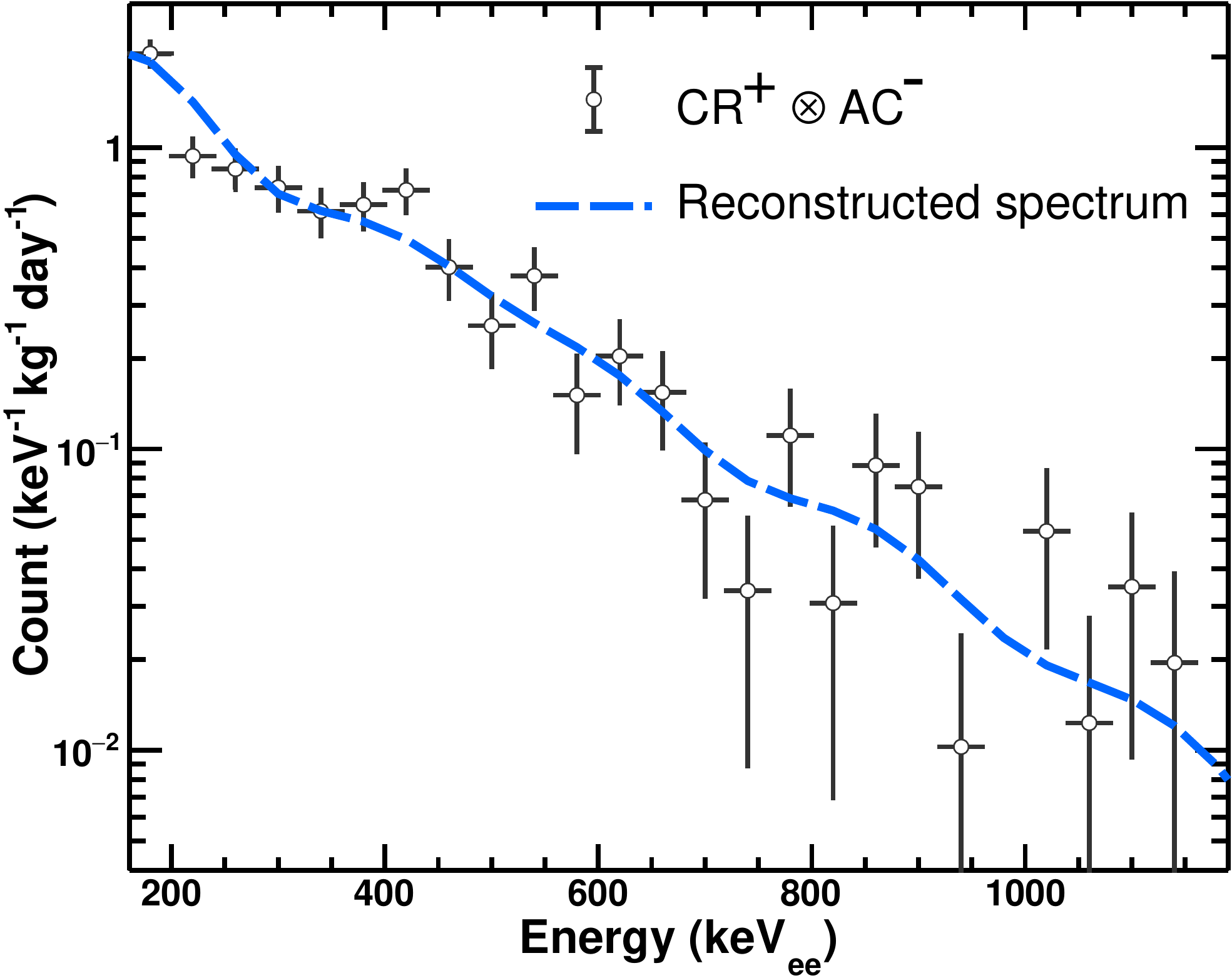} \\
{\bf (b)}\\
\includegraphics[width=8.5cm]{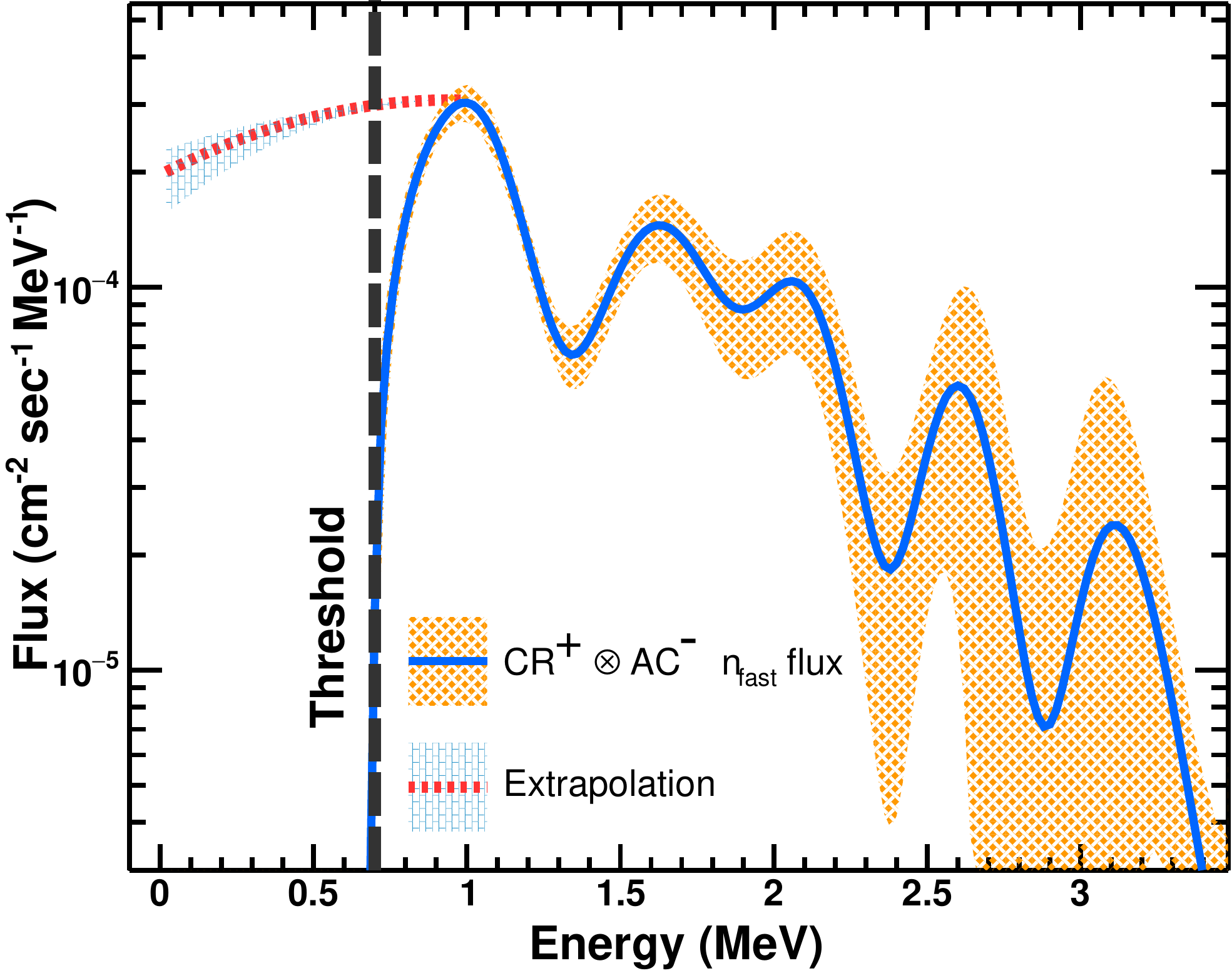} \\
{\bf (c)}\\
\includegraphics[width=8.5cm]{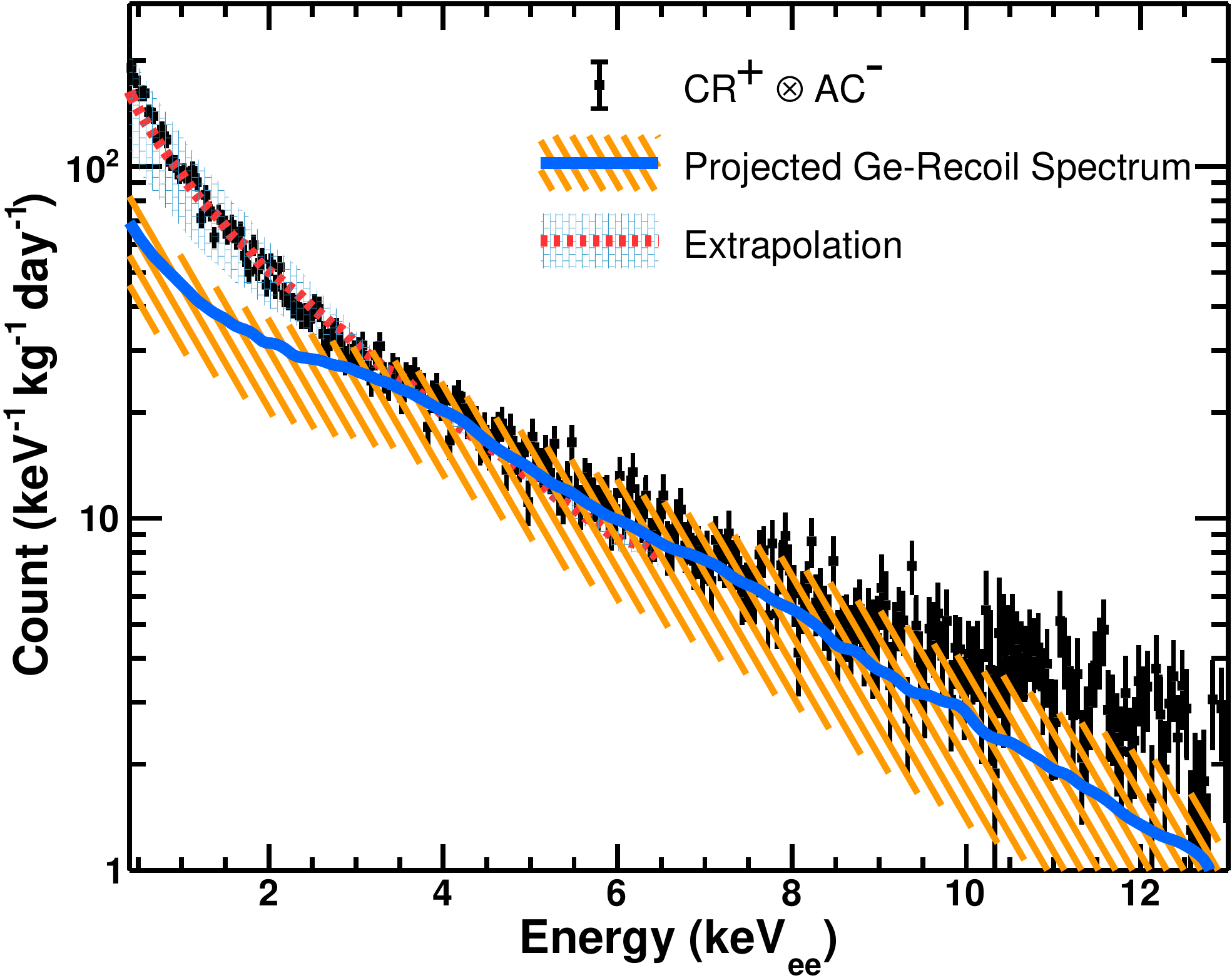}
\caption{
The sample of CR$^{+}$ $\otimes$ AC$^{-}$ --
(a) HND nuclear recoil energy spectrum, (b) unfolded neutron
flux with $\pm 1\sigma$ error as shadow area,
(c) the comparison of HPGe data and predicted Ge-recoil spectrum from
simulations with the measured neutron fluxes. Extrapolated spectra of
(b) and (c) at low energy, as fixed by neutron flux models of
Figure~\ref{fig::neut_fluxes} derived from equilibrium yield of
$^{70}$Ge(n,$\gamma$)$^{71}$Ge, are corrections to the effects due to
finite HND threshold of ${\rm 150~keV_{ee}}$.
}
\label{fig::tv_rec}
\end{center}
\end{figure}
\begin{figure}
\begin{center}
{\bf (a)}\\
\includegraphics[width=8.5cm]{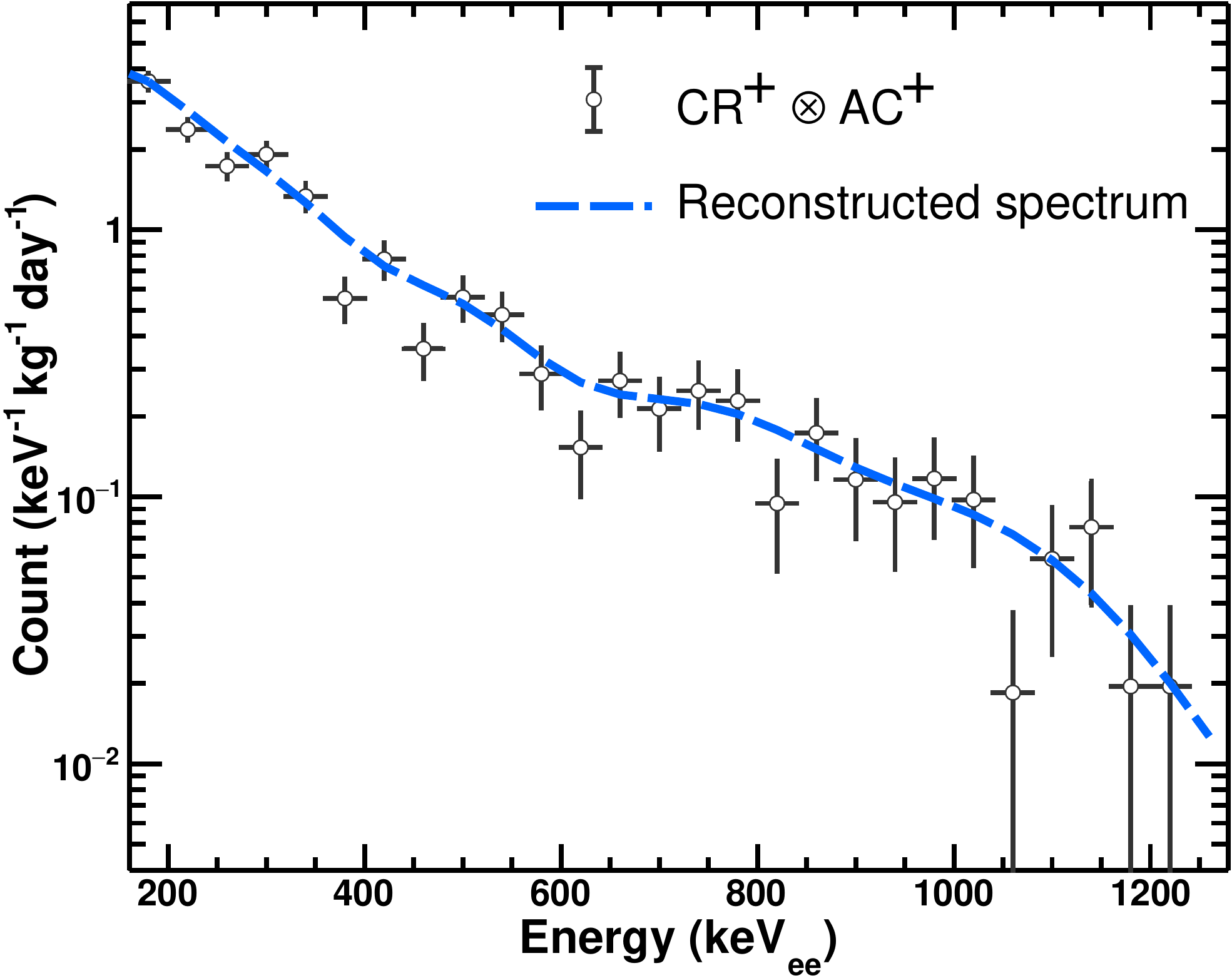} \\
{\bf (b)}\\
\includegraphics[width=8.5cm]{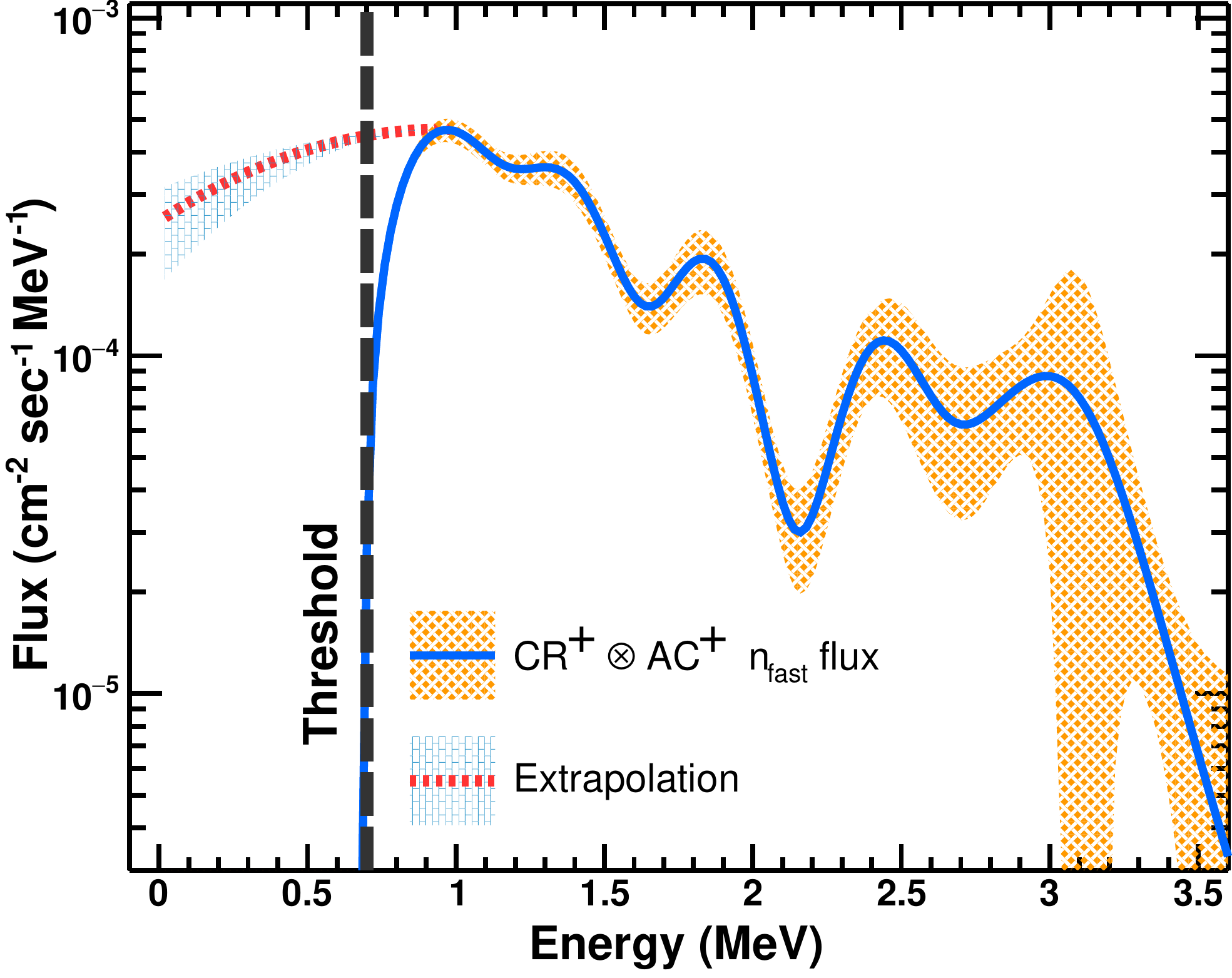} \\
{\bf (c)}\\
\includegraphics[width=8.5cm]{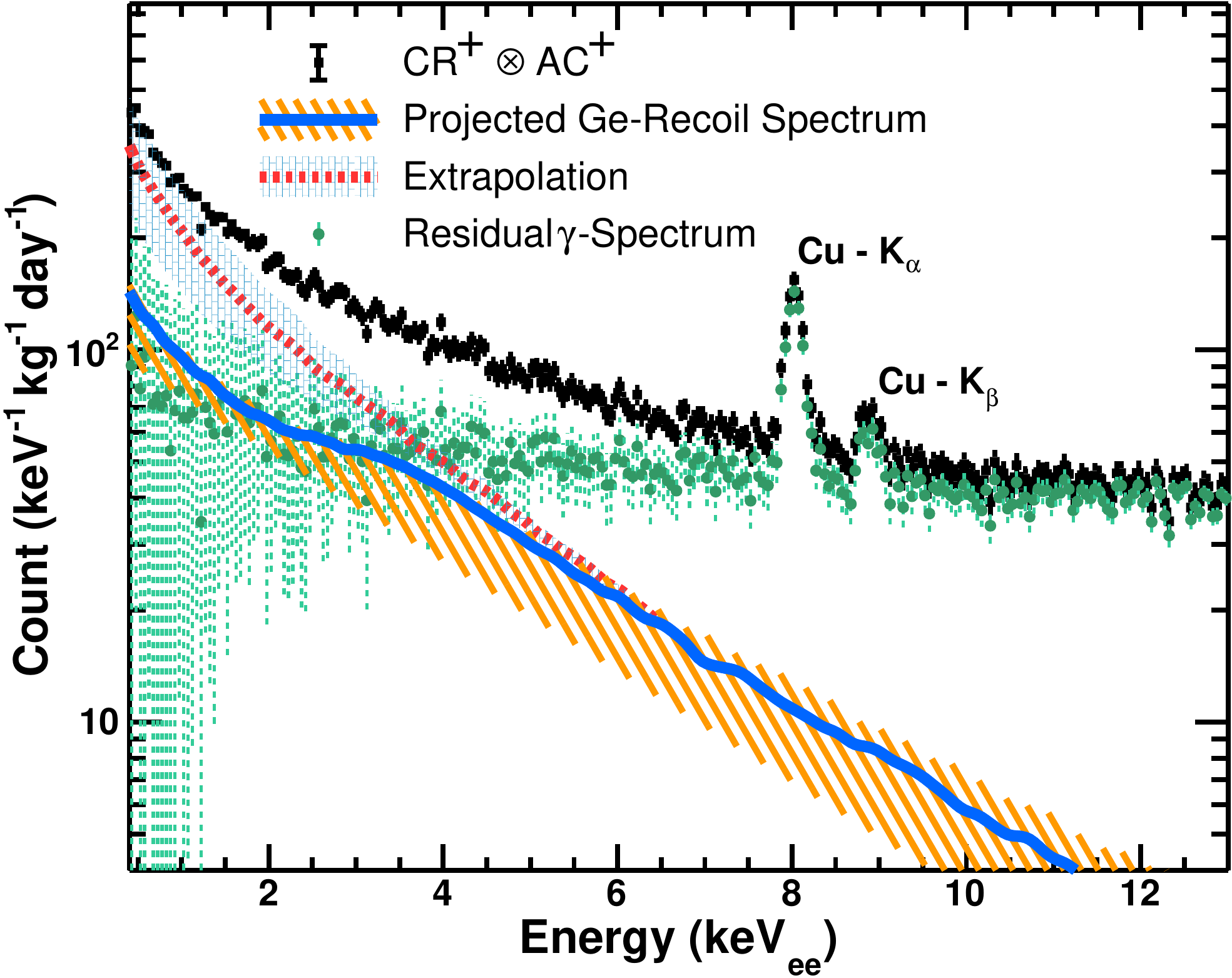}
\caption{The sample of CR$^{+}$ $\otimes$ AC$^{+}$ --
(a) HND nuclear recoil energy spectrum, (b) unfolded
neutron flux with $\pm 1\sigma$ error as shadow area,
(c) the comparison of HPGe data and predicted Ge-recoil
spectrum from simulations with the measured neutron fluxes.
Extrapolated spectra of (b) and (c) at low energy, as
fixed by neutron flux models of Figure~\ref{fig::neut_fluxes}
derived from equilibrium yield of $^{70}$Ge(n,$\gamma$)$^{71}$Ge,
are corrections to the effects due to finite HND threshold of
${\rm 150~keV_{ee}}$.
}
\label{fig::tt_rec}
\end{center}
\end{figure}
\begin{figure}
\begin{center}
{\bf (a)}\\
\includegraphics[width=8.5cm]{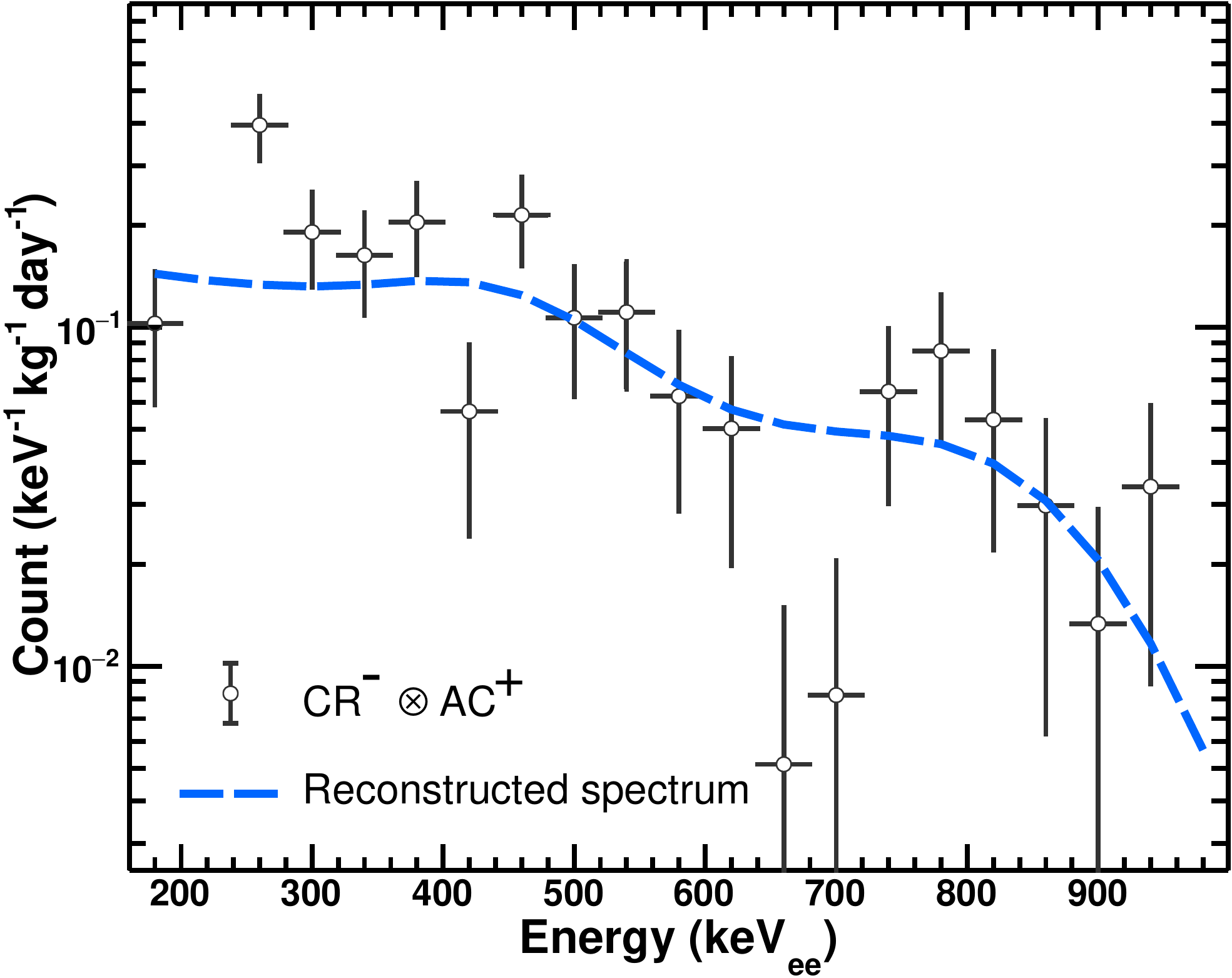}\\
{\bf (b)}\\
  \includegraphics[width=8.5cm]{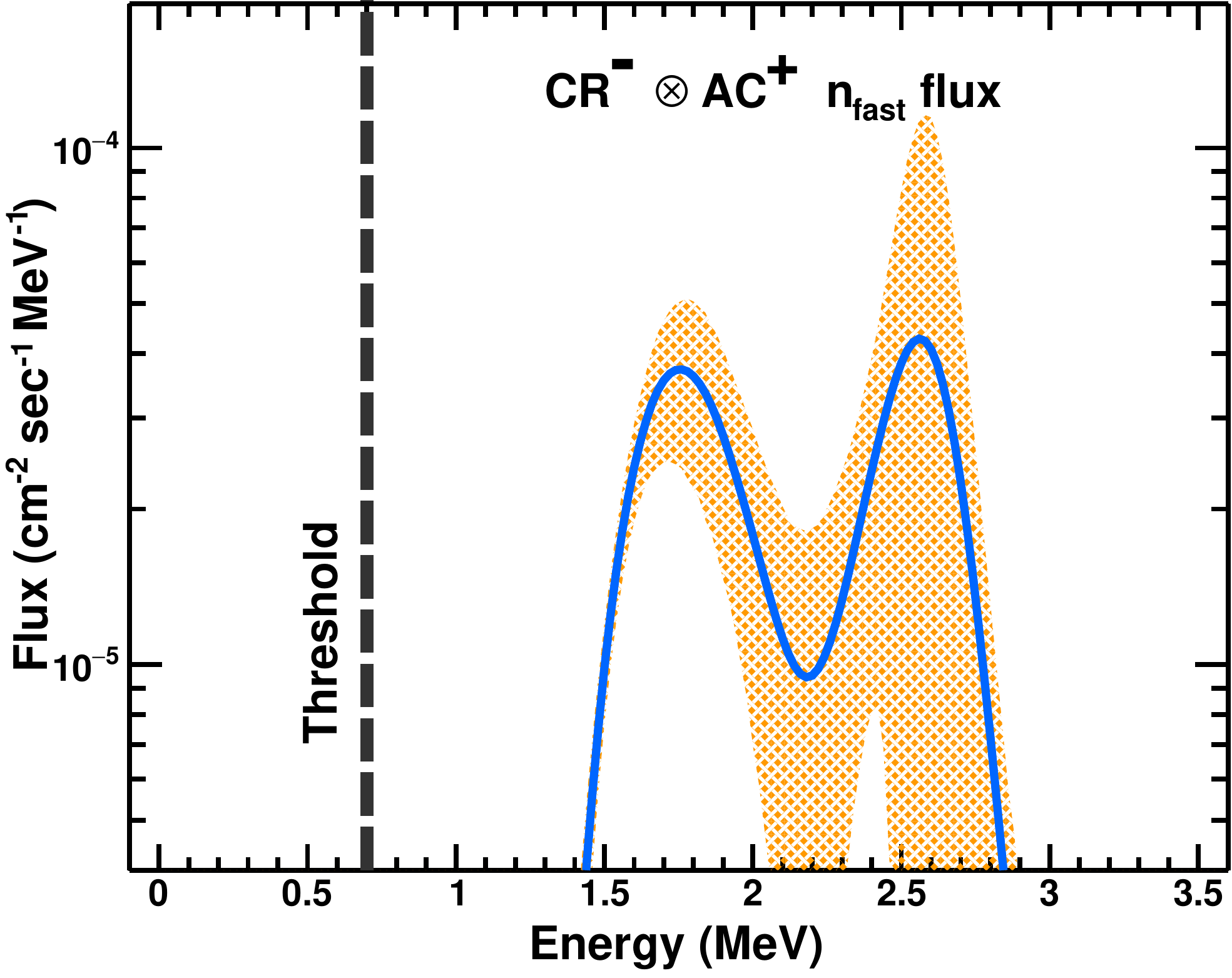} \\
{\bf (c)}\\
\includegraphics[width=8.5cm]{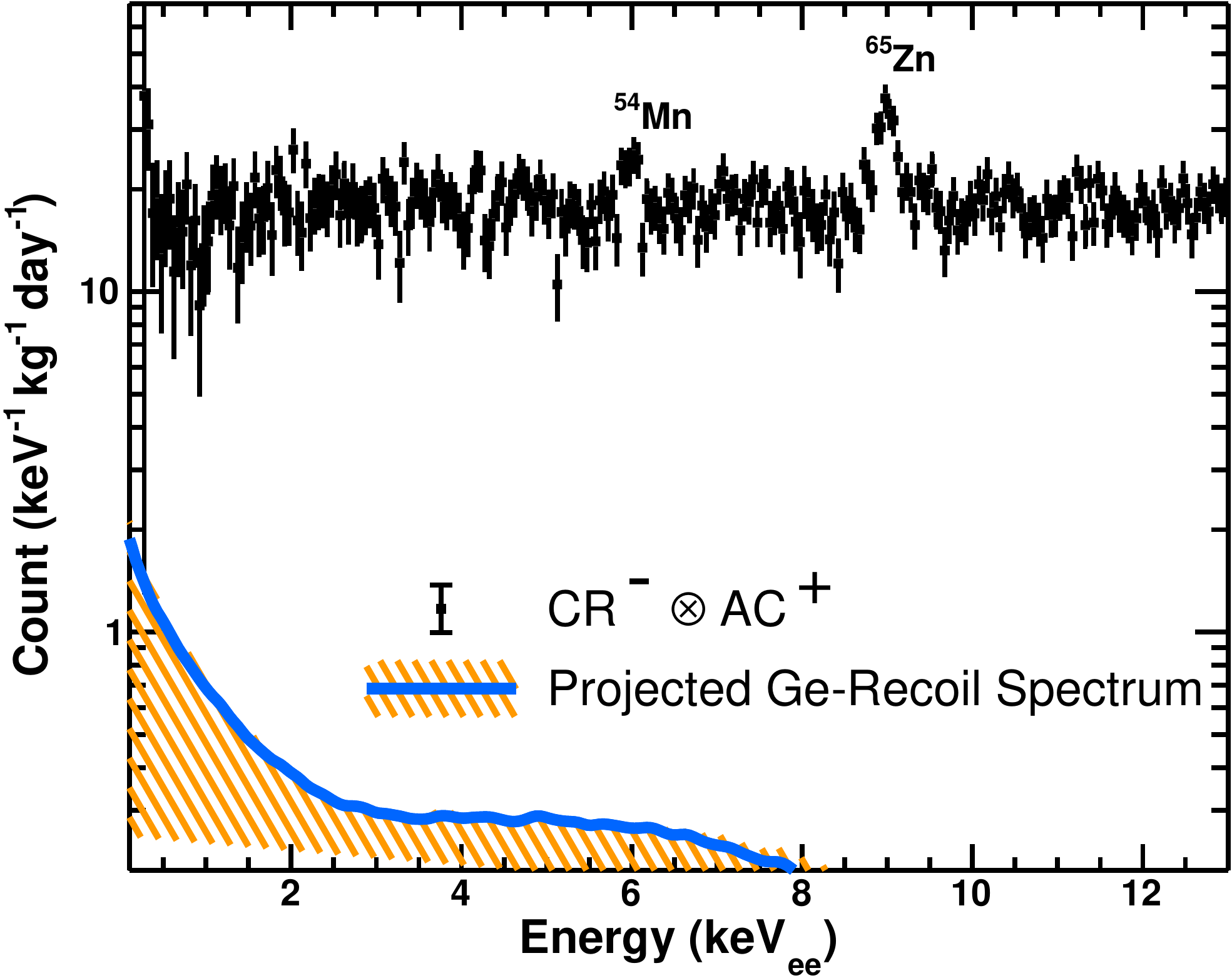}
\caption{The sample of CR$^{-}$ $\otimes$ AC$^{+}$ --
(a) HND nuclear recoil energy spectrum, (b) unfolded
neutron flux with $\pm 1\sigma$ error as shadow area,
(c) the comparison of HPGe data and predicted Ge-recoil spectrum from
simulations with the measured neutron fluxes.
}
\label{fig::vt_rec}
\end{center}
\end{figure}
\begin{figure}
\begin{center}
{\bf (a)}\\
\includegraphics[width=8.5cm]{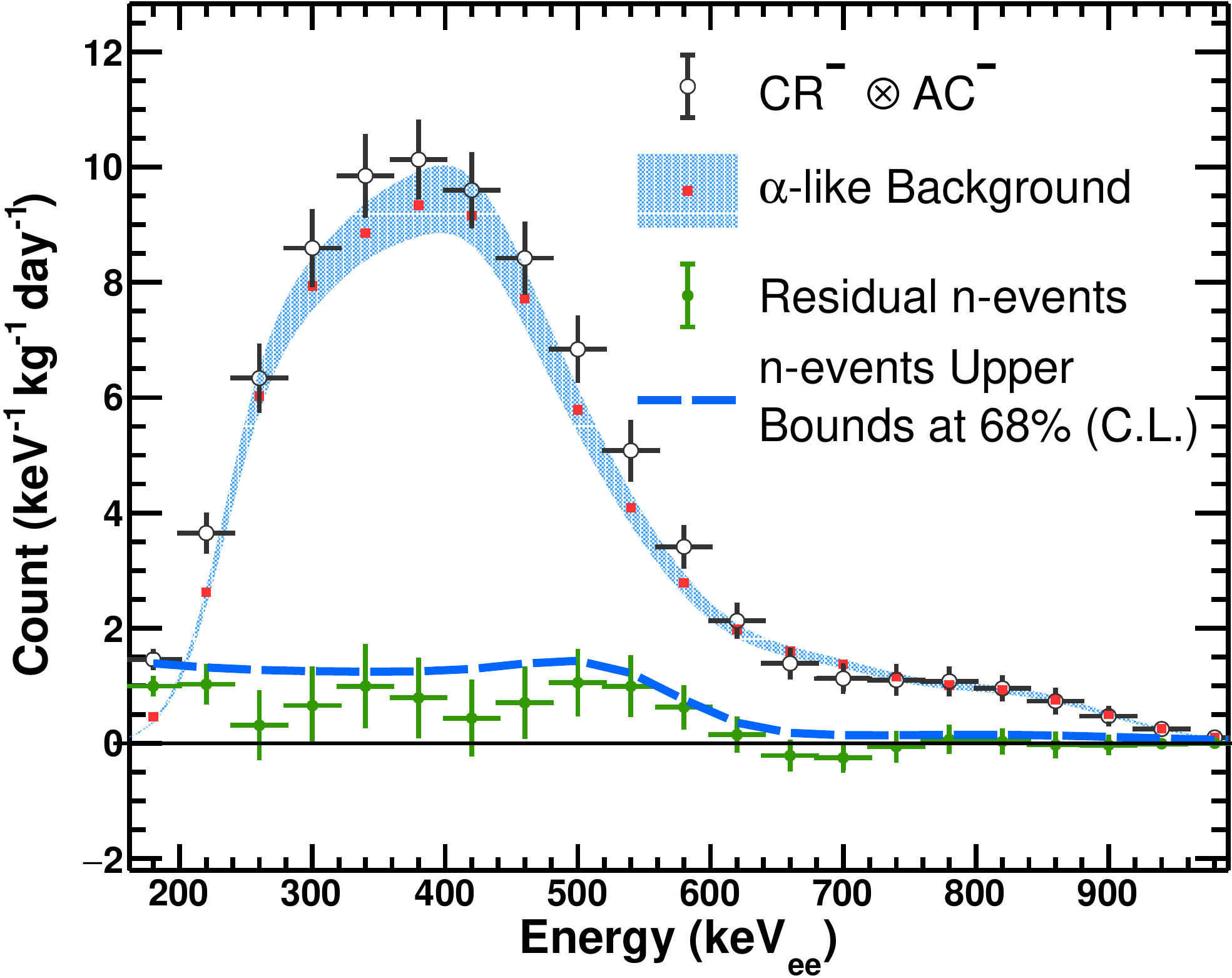}\\
{\bf (b)}\\
  \includegraphics[width=8.5cm]{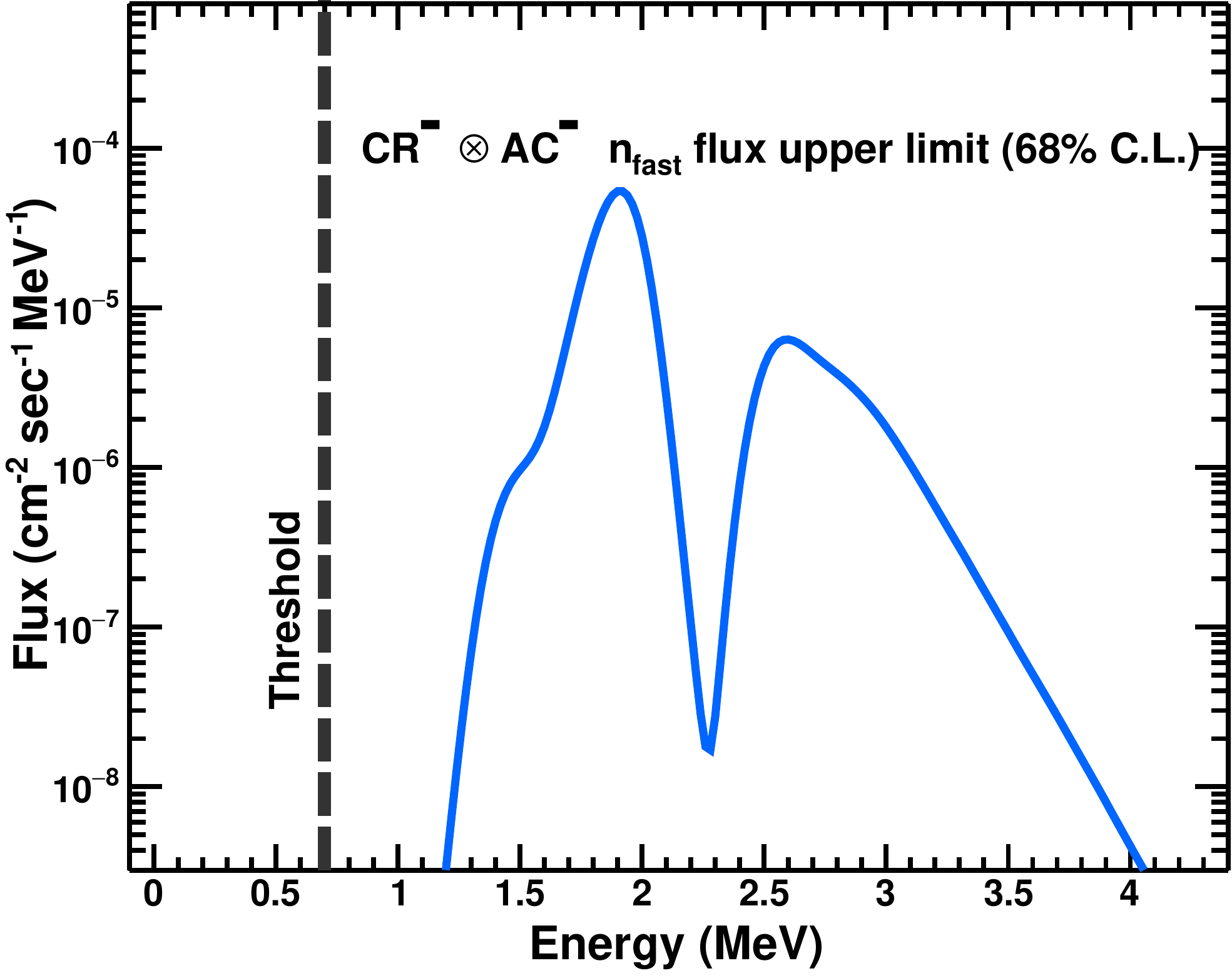} \\
{\bf (c)}\\
\includegraphics[width=8.5cm]{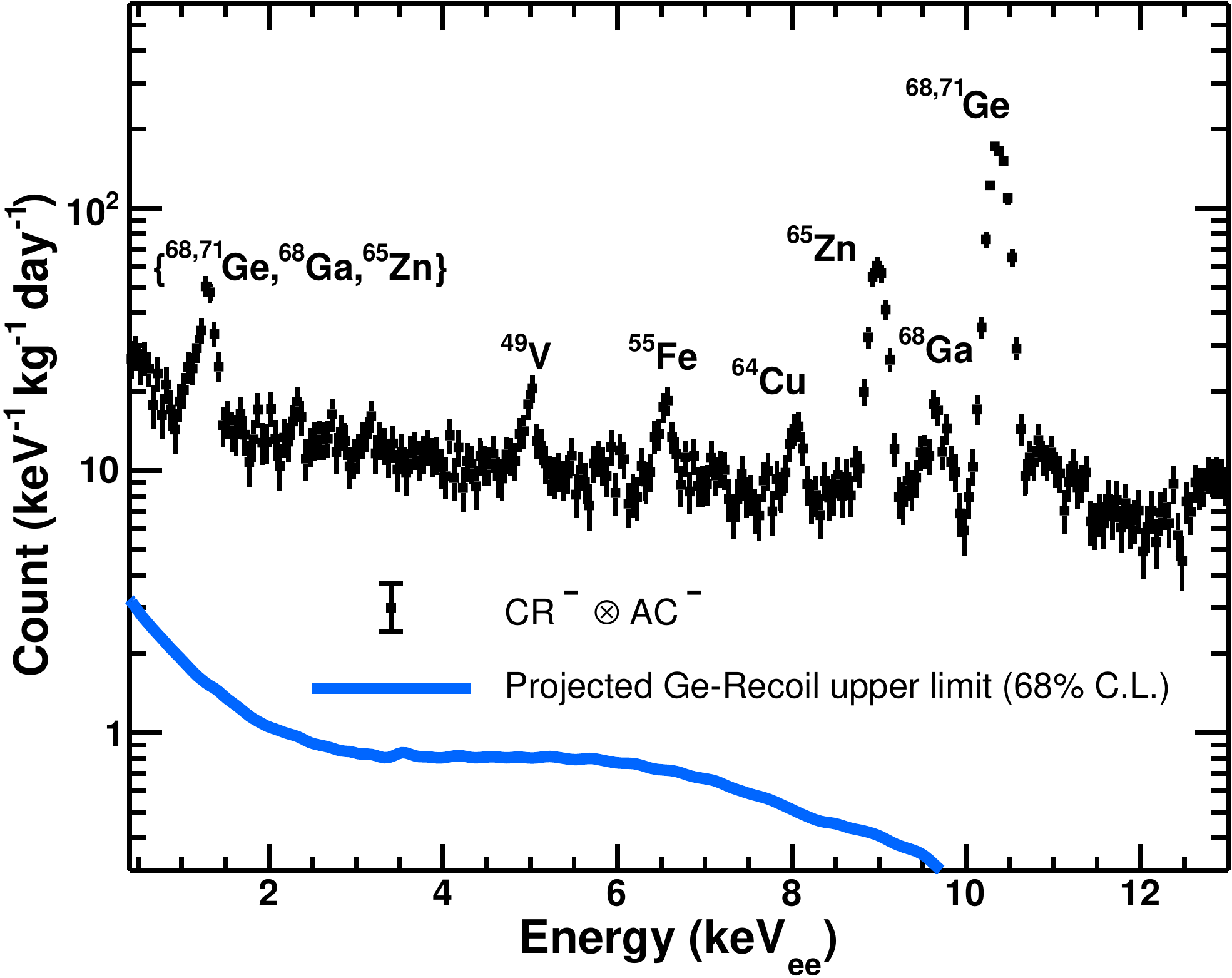}
\caption{The sample of CR$^{-}$ $\otimes$ AC$^{-}$ --
(a) energy spectra for HND nuclear recoil-like events, together
with the measured $\alpha-$ background from $\th232$ and $\u238$
decay series and the 68\% C.L. upper bound of neutron-induced
nuclear recoils, from which the upper bounds of (b) unfolded
neutron spectrum and (c) predicted Ge-recoil background in HPGe
can be derived and compared with measured data.
}
\label{fig::vv_rec}
\end{center}
\end{figure}

Once the HND spectra are measured, unfolding algorithms as
discussed in Ref.~\cite{nd_tech}, followed by a Friedman
smoothing algorithm~\cite{friedman}, are applied to produce
the corresponding fast neutron spectra. The expected nuclear
recoil background in HPGe detectors at the same location and
shielding configurations are then evaluated with full GEANT
simulation~\cite{geant} and compared with 173.5-kg-days of data
taken under identical passive and active shielding configurations
with an n-type point-contact germanium detector~\cite{qfge}.
Standard quenching function of Ge~\cite{qfge} are used to convert
nuclear recoil energy in~keV$_{\rm nr}$ into the observable energy
in electron-equivalence unit keV$_{\rm ee}$.

Results with ${\rm CR^+ \otimes AC^-}$ samples are displayed in
Figure~\ref{fig::tv_rec}, in which (a) is the recoil spectrum from
the HND liquid scintillator, (b) is the evaluated neutron spectrum
and (c) is the projected Ge recoil spectrum from the same neutron
background. The fast neutron spectrum has a threshold at ${\rm 700~keV_{nr}}$
due to HND response. The threshold effects give rise to a change of slope
of the Ge-recoil spectrum at ${\rm 4~keV_{\rm ee}}$, below which the predicted
spectrum is less than the measured one. This excesses can be corrected for
with an extrapolation to the neutron flux, the procedures and details of
which are described in Section~\ref{sect::nbkg}-C.

The same analysis procedures are applied to the $\rm{ CR^+ \otimes AC^+}$ samples,
and the results presented in Figure~\ref{fig::tt_rec} follow the same convention.
There exists a finite residual spectrum after the Ge-recoils are accounted for, as
depicted in Figure~\ref{fig::tt_rec}c. The residual events are due to Compton
scattering of cosmic-ray induced high energy ambient $\gamma$-rays, characterized
by a flat spectrum and consistent with simulations. The two peaks corresponds to
copper K$_{\alpha}$ and K$_{\beta}$ X-ray emission lines produced by the interactions
of  cosmic-ray muons with the copper support materials in the vicinity of the active
Ge crystal.

Similarly, the results of the cosmic-ray anti-coincidence samples with
$\rm{ CR^- \otimes AC^+}$ and $\rm{ CR^- \otimes AC^-}$ tags are displayed in
Figure~\ref{fig::vt_rec} and Figure~\ref{fig::vv_rec}, respectively. It can be seen
from Figure~\ref{fig::vt_rec}(c) and Figure~\ref{fig::vv_rec}(c), in both cases
that neutron-induced Ge-recoil events, which are unrelated to cosmic-rays only
constitute a minor component relative to that due to ambient $\gamma$-radioactivity.
The $\rm{ CR^- \otimes AC^-}$  events are uncorrelated with CR and AC detectors and
represent the physics candidate samples for the studies of neutrino and dark matter.
The measured ``recoil-like'' spectrum can completely be explained by internal
$\alpha$-contaminations as discussed in Section~\ref{sect::intcontam}, such that only
upper bounds for HND and HPGe as well as fast neutron spectra can be derived. The upper
limits of these spectra at 68\% C.L. are displayed in Figure~\ref{fig::vv_rec}. The peaks
in both Figure~\ref{fig::vt_rec}(c) and Figure~\ref{fig::vv_rec}(c) are due to X-rays
emissions following electron capture (EC) by the unstable isotopes, which are produced
by cosmogenic activation of long-lived isotopes inside the HPGe target.

\subsection{Complete Neutron Spectrum}

Combining both the measured thermal and fast neutron fluxes and spectra,
and adopting the neutron slowing-down theory~\cite{okay}, which is described by
a $1/E$ behavior of the epithermal region in between, the complete neutron spectrum
at KSNL can be modeled using information of the {\it in situ} measurements
of neutron capture rates.

\begin{figure}[h!]
\begin{center}
{\bf (a)}\\
\includegraphics[width=8.5cm]{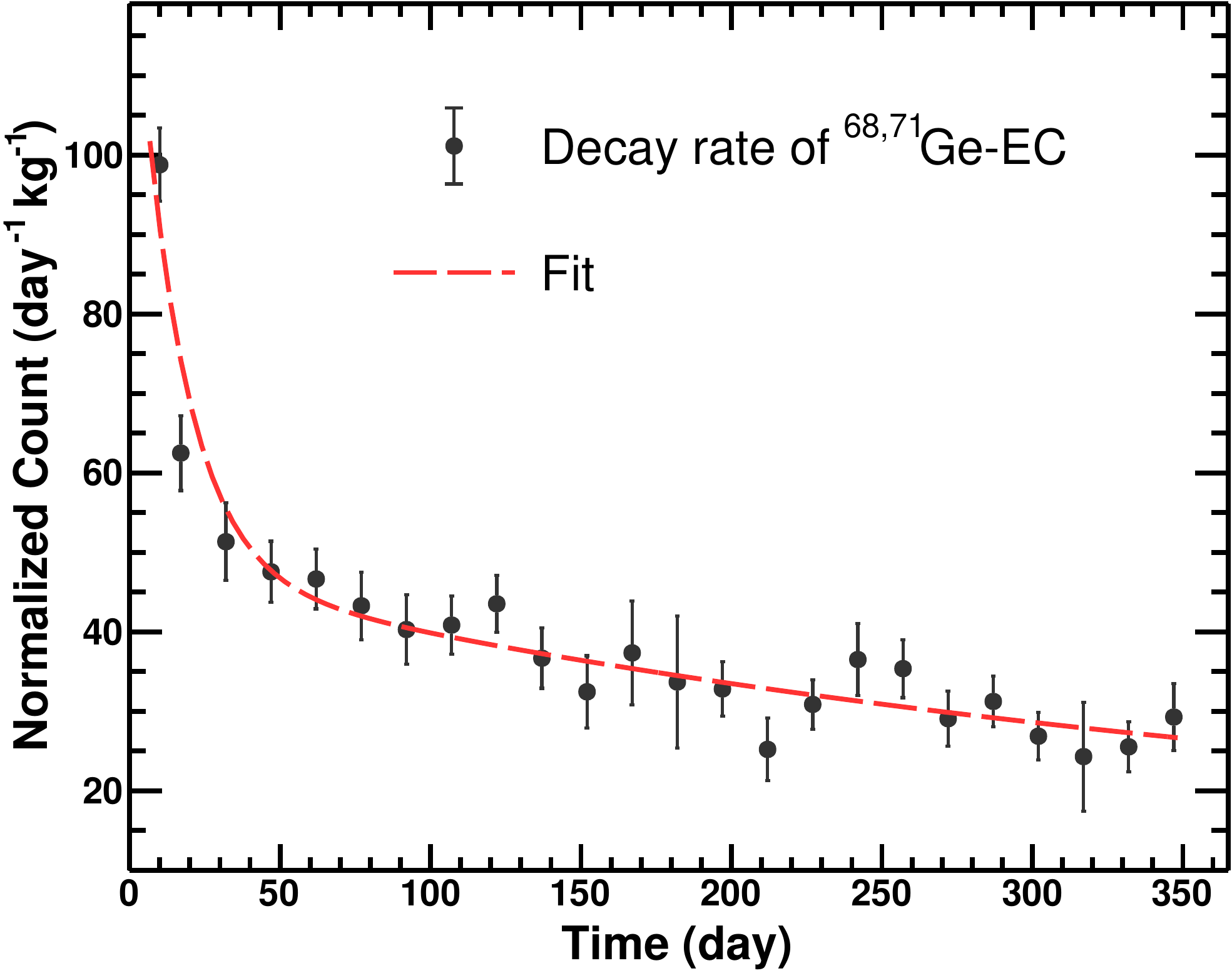} \\
{\bf (b)}\\
\includegraphics[width=8.5cm]{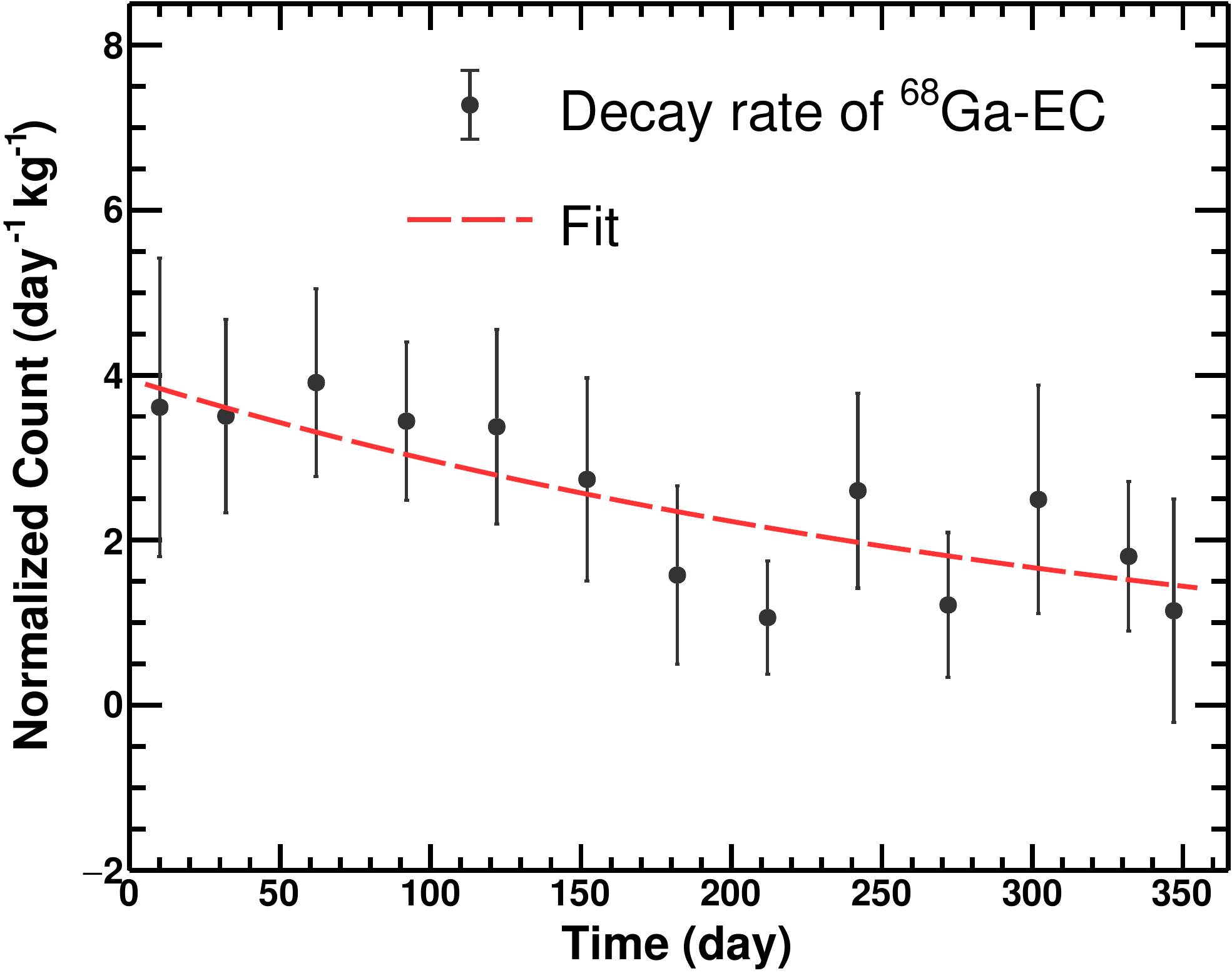}
\caption{
Time variation of characteristic K X-ray lines of (a) $^{71}$Ge and $^{68}$Ge,
and (b) $^{68}$Ga. Exponential best-fits are superimposed.
}
\label{fig::decay_cosmogenic}
\end{center}
\end{figure}

\begin{figure}[hbt]
\begin{center}
\includegraphics[width=8.5cm]{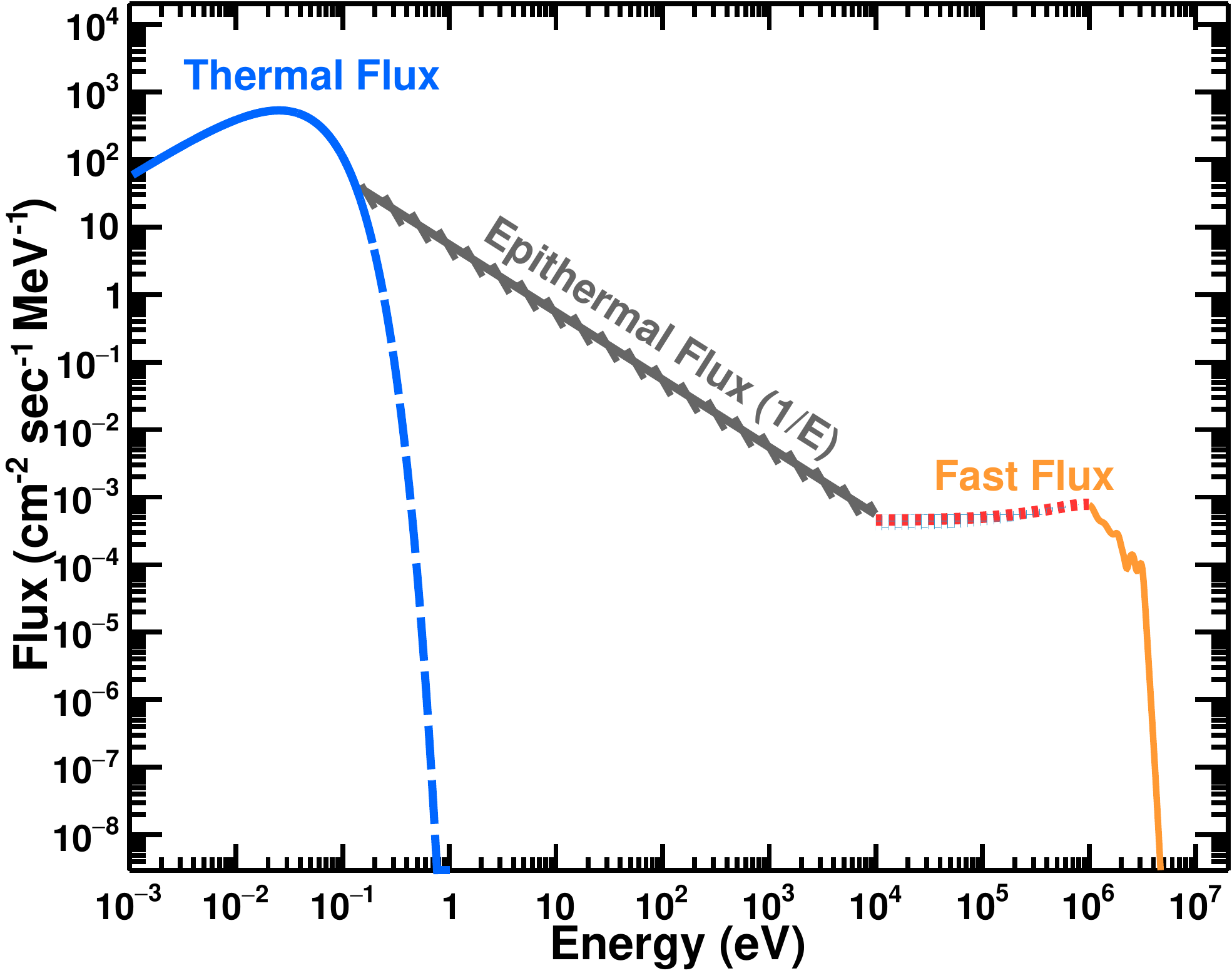}
\caption{
Neutron spectrum model at the target region of KSNL.
The total thermal and fast neutron components are based
on measurements and analysis reported in this article.
The epithermal component is from interpolation.
The cut-off at $\sim$5~MeV is a consequence of the lack of event
statistics below $\mathcal{O}( 10^{-2} )~{\rm kg^{-1} keV^{-1} day^{-1}}$
above a few ${\rm MeV_{ee}}$.
}
\label{fig::neut_fluxes}
\end{center}
\end{figure}

Figure~\ref{fig::decay_cosmogenic}a and Figure~\ref{fig::decay_cosmogenic}b
show the variations over the whole data taking period (347 days) of the X-ray
peaks at 10.37~keV$_{ee}$ and 9.66~keV$_{ee}$, respectively, following EC of
$^{71}$Ge+$^{68}$Ge and $^{68}$Ga. These isotopes are primarily produced
by neutron capture channels $^{70}$Ge(n,$\gamma$)$^{71}$Ge followed by EC
in ${\rm ^{71}Ge(e^-,\nu_e)^{71}Ga}$ and $^{70}$Ge(n,3n)$^{68}$Ge followed
by ${\rm ^{68}Ge(e^-,\nu_e)^{68}Ga}$ and  ${\rm ^{68}Ga(e^-,\nu_e)^{68}Zn}$.

\begin{table*}[]
\begin{center}
\caption{
\label{tab::estrate}
Summary of the {\it in situ} measured $^{71}$Ge/$^{68}$Ge (10.37~keV$_{ee}$) and
$^{68}$Ga (9.66~keV$_{ee}$) characteristic K X-ray line rates at KSNL $-$ for
both transient and in equilibrium components.
The equilibrium yield provide information of the {\it in situ} neutron capture yields.
The measured $^{70}$Ge(n,$\gamma$)$^{71}$Ge rates are in excellent agreement with
simulation predictions using the neutron flux model of Figure~\ref{fig::neut_fluxes}
 as input. The measured $^{70}$Ge(n,3n)$^{68}$Ge rates are consistent with zero.
There are no predictions for this channel since the threshold of $\sim$20~MeV
is above high-energy cut-off of Figure~\ref{fig::neut_fluxes}.
}
{\def\arraystretch{1.5}
\begin{tabular}{lccc}
\hline
Channel (K X-ray Lines) & \multicolumn{2}{c}{Half-Life ($\tau_{\frac{1}{2}}$)(day)} & Rate\\
Measurements & Nominal & Measured & (${\rm kg^{-1}~day^{-1}}$) \\
\hline \hline
$^{71}$Ge from Transient 10.37 keV$_{ee}$
          & 11.43
          & 10.63 $\pm$ 1.08
          & 2.70 $\pm$ 0.90\\
$^{68}$Ge from Transient 10.37 keV$_{ee}$
          & 270.95
          & 275.76 $\pm$ 9.01
          & 23.9 $\pm$ 6.4\\
$^{68}$Ge from Transient 9.66 keV$_{ee}$
          & 270.95
          & 246.74 $\pm$ 46.16
          & 2.2 $\pm$ 0.6\\
Equilibrium  9.66 keV$_{ee}$  & & & 0.05 $\pm$ 0.29 \\
~~~~=[$^{70}$Ge(n,3n)$^{68}$Ge] & & & $<$ 0.34 (68\% C.L.) \\
Equilibrium  10.37 keV$_{ee}$  & & &  \\
~~~~=[$^{70}$Ge(n,$\gamma$)$^{71}$Ge+$^{70}$Ge(n,3n)$^{68}$Ge]
     & & & 12.40 $\pm$ 3.70 \\
\hline \hline
Simulated Predictions ($\rm{kg^{-1} \times day^{-1}}$)
   & & & $^{70}$Ge(n,$\gamma$)$^{71}$Ge \\
\hline \hline
~~~~~~~~~~~~~~~~n$_{thermal}$ &  & & 8.05 $\pm$ 0.23  \\
~~~~~~~~~~~~~~~~n$_{epithermal}$ &  & & 2.18 $\pm$ 0.67 \\
~~~~~~~~~~~~~~~~n$_{fast}$ &  & & 3.67 $\pm$ 1.50 \\
~~~~~~~~~~~~~~~~Total &  & & 13.90 $\pm$ 1.65 \\
\hline
\end{tabular}}
\end{center}
\end{table*}

The decreasing intensities with time are consequences of less {\it in situ}
cosmogenic activation compared to the pre-installation activities. The measured
lifetimes are consistent with nominal values and the equilibrium levels displayed
in Table~\ref{tab::estrate}, on the other hand, provide information on the
{\it in situ} neutron capture rates of $^{70}$Ge, and hence the neutron fluxes.
It can be seen from Figure~\ref{fig::decay_cosmogenic}b that the equilibrium yield of the 9.66~keV$_{ee}$ line and hence {\it in situ} production of $^{70}$Ge(n,3n)$^{68}$Ge
are consistent with zero. Accordingly, the equilibrium yield of the 10.37~keV$_{ee}$
line is due exclusively to {\it in situ} production of $^{70}$Ge(n,$\gamma$)$^{71}$Ge.
This measured rate is used to fix the normalization of the epithermal neutron component.

\begin{figure}[hbt]
\begin{center}
\includegraphics[width=8.5cm]{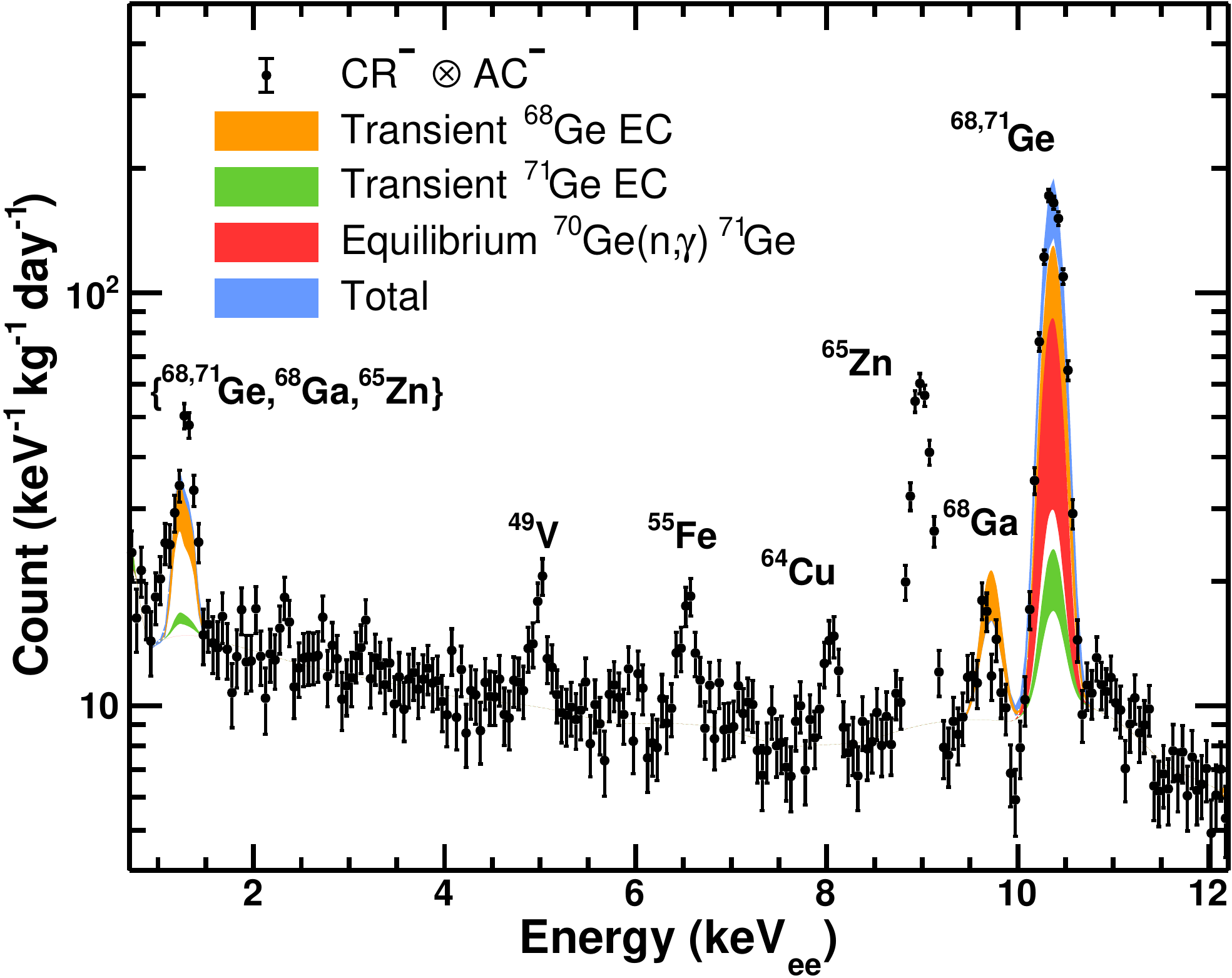}
\caption{
Measured CR$^-$ $\otimes$ AC$^-$ spectrum with HPGe at KSNL.
The various contributions to the $^{71}$Ge, $^{68}$Ge characteristic K X-ray,
L X-ray lines and the $^{68}$Ga K X-ray line, based on predictions using the
measured equilibrium neutron capture rates, are superimposed.
}
\label{fig::ge_rdecay}
\end{center}
\end{figure}

\begin{table}[hbt]
\begin{center}
\caption{
\label{tab::summary}
Summary of flux measurements of different categories of neutrons.}
{\def\arraystretch{1.5}
\begin{tabular}{lc}
\hline
Neutrons & Measured Fluxes \\
& $\Phi_n$ ${\rm (cm^{-2} s^{-1})}$ \\ \hline \hline
Thermal $-$ $0.02~{\rm eV}-1.00~{\rm eV}$ & \\
~~~~${\rm CR^{+} \otimes AC^{-}}$ & $(2.68 \pm 0.28) \times 10^{-6}$\\
~~~~${\rm CR^{+} \otimes AC^{+}}$ & $(3.00 \pm 0.29) \times 10^{-6}$\\
~~~~${\rm CR^{-} \otimes AC^{+}}$ & $(9.33 \pm 1.65) \times 10^{-7}$\\
~~~~${\rm CR^{-} \otimes AC^{-}}$ & $(2.87 \pm 0.09) \times 10^{-5}$\\
\hline
Epithermal & $\rm{\lbrace}$ \makecell[l]{$ > 4.39 \times 10^{-5}$\\
                                  $< 8.25 \times 10^{-5}$} \\
\hline
Fast $-$ $0.70~{\rm MeV}-4.00~{\rm MeV}$ & \\
~~~~${\rm CR^{+} \otimes AC^{-}}$ & $(2.35 \pm 1.60) \times 10^{-4}$\\
~~~~${\rm CR^{+} \otimes AC^{+}}$ & $(4.53 \pm 2.29) \times 10^{-4}$\\
~~~~${\rm CR^{-} \otimes AC^{+}}$ & $(1.49 \pm 5.75) \times 10^{-6}$\\
~~~~${\rm CR^{-} \otimes AC^{-}}$ & $< 3.22 \times 10^{-6}$ \\ \hline
\end{tabular}}
\end{center}
\end{table}

The complete neutron background spectrum at KSNL
is displayed in Figure~\ref{fig::neut_fluxes}. The capture rates
of $^{70}$Ge(n,3n)$^{68}$Ge and $^{70}$Ge(n,$\gamma$)$^{71}$Ge
due to the thermal, epithermal and fast neutron components
evaluated by full GEANT simulations~\cite{geant} are listed in
Table~\ref{tab::estrate}, the sum of which is in excellent
agreement with the measured rates. The consistency is illustrated
in the measured CR$^-$ $\otimes$ AC$^-$ HPGe spectra of
Figure~\ref{fig::ge_rdecay} in which the different components of
the Ge K X-ray lines are shown. Their total fluxes under different
tags are given in Table~\ref{tab::summary}.
The high-energy cut-off at $\sim$5~MeV in Figure~\ref{fig::neut_fluxes}
is a consequence of the lack of event statistics below
$\mathcal{O}( 10^{-2} )~{\rm kg^{-1} keV^{-1} day^{-1}}$ for proton recoils at
energy above few ${\rm MeV_{ee}}$. This, however, would not affect background studies and understanding of the HPGe experiments at KSNL, since the background is dominated
by the lower energy background neutron which has much higher intensity.

Once the complete neutron background is modeled,  the cut-off effects
of the fast neutron spectra around 700~keV in Figure~\ref{fig::tv_rec}(b)
and Figure~\ref{fig::tt_rec}(b) due to the HND threshold at 150~keV$_{ee}$
can be corrected by extrapolation to lower energy. The corrected
Ge-recoil spectra with the additional neutrons taken into account are displayed
in Figure~\ref{fig::tv_rec}(c) and Figure~\ref{fig::tt_rec}(c). The corrected
CR$^+$ $\otimes$ AC$^-$ Ge-recoil spectrum provides $>$99\% match
to the measured data, confirming the expected physical picture where
CR$^+$ $\otimes$ AC$^-$ samples are dominated by nuclear recoil events due to
interactions of cosmic-ray induced fast neutrons. Similarly, the corrected
CR$^+$ $\otimes$ AC$^+$ Ge-recoil spectrum has the expected exponential decrease
with energy. Once accounted for, the residual cosmic-induced  $\gamma$-background
is flat down to sub-keV,  also expected from Compton scattering of high energy
$\gamma$-rays. The consistencies of these independent measurements serve as
non-trivial cross-checks on the validity of neutron flux measurements as well as
the experimental approaches and analysis procedures reported in this work.

\section{Summary and Prospects}

We report in this article {\it in situ} measurements of neutron-induced
background at KSNL with a HND under identical active and passive shielding
configurations during the neutrino physics measurements. The different
components of neutron fluxes thus derived are summarized in
Table~\ref{tab::summary}, and the neutron spectrum is depicted in
Figure~\ref{fig::neut_fluxes}. The derived neutron spectrum provides excellent
agreement with the cosmic-ray neutron-induced Ge-recoil spectra as shown in
Figure~\ref{fig::tv_rec}(c) and Figure~\ref{fig::tt_rec}(c), thereby providing
strong support to the validity of the results as well as the experimental
approaches and analysis procedures.

It was demonstrated that elastic nuclear recoil events due to cosmic-ray induced
high energy neutrons contribute almost exclusively to the $\rm{ CR^{+} \otimes AC^{-}}$
channel below 12~keV$_{ee}$,  and are major components of the $\rm{ CR^{+} \otimes AC^{+}}$
channel, dominating over $\gamma$-induced background below 4~keV$_{ee}$. On the other hand,
contributions of cosmic-uncorrelated neutrons to the background  are minor in
$\rm{ CR^{-} \otimes AC^{+}}$ and unobservable in $\rm{ CR^{-} \otimes AC^{-}}$.
In particular, the dominant background to the studies of neutrinos, WIMP dark matter
and axions with $\rm{ CR^{-} \otimes AC^{-}}$ selection at KSNL are ambient
$\gamma$-radioactivity and intrinsic cosmogenic activation.  Contributions of neutrons
from ambient radioactivity and reactor operation are negligible, a feature consistent
with expectations from full GEANT simulations. The HND detector concept and analysis
procedures can be applicable to characterize neutron background in other rare-event
experiments, in both surface and underground laboratories. In particular, the equilibrium
levels of the X-ray peaks in HPGe detectors can be used to measure {\it in situ} background
neutron fluxes. This technique can be extended to other Ge-based underground WIMP-search
experiments\cite{cogent,cdex,cdms}.

\section{Acknowledgments}

This work is supported by Contract No. 114F374 under Scientific and Technological
Research Council of Turkey (TUBITAK), the Academia Sinica Principal Investigator Award,
Contract No. 104-2112-M-001-038-MY3 from the Ministry of Science and Technology, Taiwan
and Contract No. 2017-ECP2 from the National Center of Theoretical Sciences, Taiwan.
M.K. Singh thanks the University Grants Commission (UGC), Govt. of India, for the
funding through UGC D.S. Kothari Postdoctoral Fellowship (DSKPDF) scheme (No. F. 4-
2/2006 (BSR)/PH/15-16/0098).


\end{document}